\documentclass[aps,
               pra,
               showpacs,
               amssymb,
               nofootinbib,
               superscriptaddress,
               twocolumn,
               10pt,
               longbibliography]{revtex4-2}
\usepackage[ansinew]{inputenc}
\usepackage{bbold}
\usepackage{bbm}
\usepackage{bm}
\usepackage{amsbsy}
\usepackage{amsthm}
\usepackage{amssymb}
\usepackage{amsfonts}
\usepackage{amssymb}
\usepackage{amsmath, mathtools}
\usepackage{dsfont}
\usepackage{graphicx}
\usepackage{epsfig}
\usepackage{epstopdf}
\usepackage{dsfont}
\usepackage{mathrsfs} 
\usepackage{multibib}
\usepackage[dvipsnames]{xcolor}
\usepackage[colorlinks]{hyperref}
\usepackage{subfigure}
\makeatletter
\newcommand\org@hypertarget{}
\let\org@hypertarget\hypertarget
\renewcommand\hypertarget[2]{%
  \Hy@raisedlink{\org@hypertarget{#1}{}}#2%
  }
\makeatother
\usepackage[figure,table]{hypcap}

\usepackage{MnSymbol}
\usepackage{enumerate}
\usepackage{float}
\usepackage{comment}
\usepackage{braket}
\usepackage{booktabs}
\usepackage{orcidlink}
\usepackage{tcolorbox}
\makeatletter
\newcommand{\providetcbcountername}[1]{%
  \@ifundefined{c@tcb@cnt@#1}{%
    --undefined--%
  }{%
    tcb@cnt@#1%
  }
}

\newcommand{\settcbcounter}[2]{%
  \@ifundefined{c@tcb@cnt@#1}{%
    \GenericError{Error}{counter name #1 is no tcb counter }{}{}%
  }{%
    \setcounter{tcb@cnt@#1}{#2}%
   }%
}%
\tcbuselibrary{theorems}
\newtcolorbox[blend into=tables]{smallboxtable}[3][]
{colback=mycolor!5, colframe=mycolor, float=ht, width=0.48\textwidth, lower separated=false, blend before title=colon hang,
title={#2}, label=#3 ,#1}
\newtcbtheorem{Definitions}{Definition}%
{colback=mycolor!5,colframe=mycolor,fonttitle=\bfseries,
left=4pt,
right=4pt,
top=5pt,
bottom=5pt}{defi}
\newtcbtheorem{Theorems}{Theorem}%
{colback=mycolor!5,colframe=mycolor,fonttitle=\bfseries,
left=4pt,
right=4pt,
top=5pt,
bottom=5pt}{theorem}
\definecolor{mycolor}{rgb}{0.122, 0.435, 0.698}
\usepackage{makecell}

\hypersetup{
	bookmarksnumbered,
	pdfstartview={FitH},
	citecolor={darkgreen},
	linkcolor={darkred},
	urlcolor={darkblue},
	pdfpagemode={UseOutlines}}
\definecolor{darkgreen}{RGB}{50,190,50}
\definecolor{darkblue}{RGB}{0,0,190}
\definecolor{darkred}{RGB}{238,0,0}
\definecolor{mycolor}{rgb}{0.122, 0.435, 0.698}
\usepackage{soul}

\newcommand{\scpr}[2]{\ensuremath{\left\langle\right.\hspace*{-1pt} #1 \hspace*{-1pt}\left|\right.\hspace*{-1pt} #2 \hspace*{-1pt}\left.\right\rangle}}
\newcommand{\nl}{\ensuremath{\hspace*{-0.5pt}}}
\newcommand{\nr}{\ensuremath{\hspace*{0.5pt}}}

\newcommand{\subtiny}[3]{\ensuremath{_{\hspace{#1 pt}\protect\raisebox{#2 pt}{\tiny{$ #3$}}}}}
\newcommand{\suptiny}[3]{\ensuremath{^{\hspace{#1 pt}\protect\raisebox{#2 pt}{\tiny{$ #3$}}}}}
\newcommand{\expval}[1]{\ensuremath{\left\langle\right.\hspace*{-1pt} #1 \hspace*{-1pt}\left.\right\rangle}}

\newcommand{\tr}{\textnormal{Tr}}

\DeclareMathOperator{\diag}{diag}

\ExplSyntaxOn
    \seq_new:N \g_other_authors_seq
    \seq_gset_from_clist:Nn \g_other_authors_seq
      {
        Simon~Morelli,
        Kl{\'a}ra~Baksov{\'a},
        Phila~Rembold
      }
    \seq_gshuffle:N \g_other_authors_seq
    \NewDocumentCommand{\printotherauthors}{}{
      \seq_use:Nn \g_other_authors_seq {,~}
    }
\ExplSyntaxOff

\newtheorem{theorem}{Theorem}
\newtheorem{lemma}{Lemma}

\renewcommand{\thesection}{\Roman{section}}
\renewcommand{\thesubsection}{\Roman{section}.\Alph{subsection}}
\renewcommand{\thesubsubsection}{\Roman{section}.\Alph{subsection}.\arabic{subsubsection}}
\makeatletter
\renewcommand{\p@subsection}{}
\renewcommand{\p@subsubsection}{}
\makeatother

\begin{document}
\title{Certification of high-dimensional entanglement in continuous-variable systems}
\author{Ida Mishra\, \orcidlink{0009-0003-9543-4775}}
\email{ida.mishra@tuwien.ac.at}
\affiliation{Technische Universit{\"a}t Wien, Atominstitut \& Vienna Center for Quantum Science and Technology (VCQ),  Stadionallee 2, 1020 Vienna, Austria}
\author{Simon Morelli\,\orcidlink{0000-0002-7588-2701}}
\affiliation{Technische Universit{\"a}t Wien, Atominstitut \& Vienna Center for Quantum Science and Technology (VCQ),  Stadionallee 2, 1020 Vienna, Austria}
\author{Kl{\'a}ra Baksov{\'a}\,\orcidlink{0009-0009-8944-6044}}
\affiliation{Charles University, Faculty of Mathematics and Physics, Ke Karlovu 3, 121 16 Prague, Czech Republic}
\affiliation{Technische Universit{\"a}t Wien, Atominstitut \& Vienna Center for Quantum Science and Technology (VCQ),  Stadionallee 2, 1020 Vienna, Austria}
\author{Phila Rembold\,\orcidlink{0000-0003-1405-730X}}
\affiliation{Technische Universit{\"a}t Wien, Atominstitut \& Vienna Center for Quantum Science and Technology (VCQ),  Stadionallee 2, 1020 Vienna, Austria}
\author{Nicolai Friis\,\orcidlink{0000-0003-1950-8640}}
\email{nicolai.friis@tuwien.ac.at}
\affiliation{Technische Universit{\"a}t Wien, Atominstitut \& Vienna Center for Quantum Science and Technology (VCQ),  Stadionallee 2, 1020 Vienna, Austria}

\date{\today}

\begin{abstract}
    The dimensionality of entanglement in high-dimensional quantum systems is quantified by the Schmidt number, whose large value signals potential advantages for quantum communication and quantum computation. A commonly used approach to practical Schmidt-number detection in finite-dimensional systems is based on the estimation of fidelities to pure states with large entanglement dimensionality or lower bounds to these fidelities. However, for mode-entangled continuous-variable systems, the corresponding measurements are typically difficult to realize. Here, we introduce and assess fidelity bounds for Schmidt-number detection based on practically more accessible measurements of the first and second statistical moments of arbitrary continuous-variable quantum states. We provide analytical formulas that allow evaluating these bounds and test them for families of Gaussian and non-Gaussian states, including two-mode squeezed thermal states and photon-added displaced two-mode squeezed vacuum states. We compare our bounds to other methods for the certification of high-dimensional entanglement based on the same measurements. 
\end{abstract}

\date{\today}

\maketitle


\section{Introduction}\label{sec:intro}

{\noindent}Entanglement is a form of correlation intrinsic to quantum systems with no classical counterpart. It is a crucial resource for many quantum information-processing tasks, from quantum teleportation~\cite{BennettBrassardPopescuSchumacherSmolinWooters1996}, dense coding~\cite[Sec.~2.3]{NielsenChuang2010}, and entanglement-based quantum key distribution ~\cite{Ekert1991}, to quantum computation~\cite{JozsaLinden2003}, and quantum metrology~\cite{GiovannettiLloydMaccone2006}. 
While the simplest quantum system exhibiting entanglement comprises a pair of qubits{\textemdash}two-dimensional quantum systems{\textemdash}many interesting phenomena, e.g., bound entanglement~\cite{Horodecki1997}, only occur in higher dimensions. 
Furthermore, high-dimensional (HD) quantum systems offer potential advantages, such as entanglement-enhanced imaging~\cite{NdaganoDefienneLyonsStarshynovVillaTisaFaccio2020, DefienneNdaganoLyonsFaccio2021, CameronCourmeVernierePandyaFaccioDefienne2024}, noise-robust entanglement distribution~\cite{Ecker-Huber2019, ZhuTylerHerreraValenciaMalikLeach2021}, and loophole-free Bell tests~\cite{SalavrakosAugusiakTuraWittekAcinPironio2017, VertesiPironioBrunner2010}. 
Recent applications have been demonstrated in quantum computing~\cite{WangHuSandersKais2020, KiktenkoNikolaevaFedorov2025}, but HD entanglement plays a crucial role in quantum communication~\cite{CozzolinoDaLioBaccoOxenloewe2019} leading to record capacities~\cite{IslamEtAl2017, MirhosseiniEtAl2015} and increased secure key rates~\cite{GroeblacherJenneweinVaziriWeihsZeilinger2006, HuberPawlowski2013, DodaHuberMurtaPivoluskaPleschVlachou2021}.
Consequently, there are rapid developments~\cite{MalikEtAl2026} in the theory describing the generation, distribution, and quantification of HD entanglement (see~\cite{GuehneToth2009,EltschkaSiewert2014,FriisVitaglianoMalikHuber2019,ErhardKrennZeilinger2020} or~\cite[Chapter~17]{BertlmannFriis2023} for reviews).

A central problem connected to the use of this resource is the certification of HD entanglement from practically implementable measurements. For pairs of finite-dimensional quantum systems, measurements in two or more complementary bases have been shown to be sufficient for obtaining lower bounds on fidelities with suitable pure target states, which in turn provide lower bounds~\cite{BavarescoEtAl2018} on the Schmidt number (SN)~\cite{TerhalHorodecki2000}. When these bases are mutually unbiased, this method also provides a lower bound on the entanglement of formation~\cite{BavarescoEtAl2018}, and has been successfully employed in practice~\cite{Ecker-Huber2019, HerreraValenciaSrivastavPivoluskaHuberFriisMcCutcheonMalik2020, ChangSarihanLiEtAl2026}. Meanwhile, measurements in mutually unbiased bases (MUBs) can also be used in different ways: firstly, to directly obtain lower bounds on the SN~\cite{MorelliHuberTavakoli2023,TavakoliMorelli2024}, a method that has in turn been generalized to measurements in arbitrary bases~\cite{LiHuberFriis2025}, and secondly, to obtain lower bounds on the distillable entanglement via entropic uncertainty relations~\cite{DevetakWinter2005,BertaChristandlColbeckRenesRenner2010}. 

Here, we focus on the certification of HD entanglement in systems that, arguably, have the potential for the highest entanglement dimensionality: infinite-dimensional quantum systems. More specifically, we focus on pairs of continuous-variable (CV) quantum systems. Motivated by the success of MUBs for the certification of HD entanglement in finite dimensions, we investigate how well HD entanglement can be certified in infinite dimensions using measurements of observables that are both complementary and practically easily accessible in CV systems in a similar fashion as MUBs are in finite dimensions. 

To this end, we derive new SN bounds based only on measurements of the first and second statistical moments of complementary quadrature operators. 
Our method leverages recent results~\cite{MeleMeleBittelEisertGiovannettiLamiLeoneOliviero2025} on bounding the trace distance of arbitrary states to pure Gaussian states in combination with Fuchs{\textendash}van~de~Graaf inequalities~\cite{FuchsVanDeGraaf1999} to obtain lower bounds on the fidelity with pure Gaussian target states. Finally, we apply bounds on the maximal fidelity that any states with SN~$k$ can achieve with pure states (see~\cite[Supplementary Material, Sec. C]{FicklerEtAl2014},~\cite{BavarescoEtAl2018}, or~\cite[Theorem 17.3]{BertlmannFriis2023}) to arrive at a family of fidelity-based SN witnesses that can be evaluated using only the covariance matrix (CM) of the otherwise unknown state.

By design, our method works well for states whose CM is close to that of any pure entangled two-mode Gaussian state but performs less well for states whose CM is associated to a Gaussian state with higher mixedness. 
We therefore test our method for two families of states: First, we consider two-mode squeezed thermal (TMST) states, parameterized by the initial temperature and the squeezing strength. 
Second, we study photon-added displaced~(PAD) two-mode squeezed vacuum (TMSV) states, parameterized by the initial squeezing and the displacement before the photon addition. While the former family consists of only Gaussian states, it allows us to illustrate the interplay of increased mixedness and squeezing on the detection of the SN. In contrast, the second family contains only pure but non-Gaussian states, which, given proper normalization, converge to a pure and Gaussian displaced TMSV state in the limit of infinite displacement.  

Thus, we cover pure and mixed as well as Gaussian and non-Gaussian states. Moreover, we compare our method to two others that only require the state's CM: The first one bounds the SN via the entanglement of formation~(EoF)~\cite{BennettDiVincenzoSmolinWootters1996}. The EoF can be calculated analytically from the CM for all two-mode Gaussian states~\cite{AkbariKourbolaghAlijanzadehBoura2015}. In turn, the EoF of a Gaussian state bounds the EoF of any other state with the same CM from below~\cite{WolfGiedkeCirac2006}. 
The second method is based on linear Gaussian SN witnesses~\cite{ShahandehSperlingVogel2013}, i.e., linear operators that are at most quadratic polynomials in the mode operators of the involved modes, and whose expectation values are hence functions of the first and second moments only. 
At the same time, the expectation value of these Gaussian SN witnesses with any SN-$k$ state can be bounded by solving the corresponding SN-eigenvalue equations~\cite{SperlingVogel2011}, such that a violation of this bound indicates a SN larger than~$k$. 

We find that, for the examples considered, the linear Gaussian SN witnesses~\cite{ShahandehSperlingVogel2013} perform best in most settings in detecting the SN, with no particular ordering among the other methods in terms of performance. However, while the linear Gaussian witnesses only determine a lower bound on the SN, the other methods give additional information on the fidelity with Gaussian pure states and the EoF, respectively. Generally, which method ultimately performs best for states outside our examples is not revealed by our work. 
 
In this work, we do not compare approaches based on more elaborate measurements, e.g., SN witnesses for CV systems based directly on covariances of infinite-dimensional Bloch operators~\cite{LiuGuoFadelHeHuberVitagliano2025}, or discrete-variable SN witnesses based either on measurements in MUBs~\cite{BavarescoEtAl2018,MorelliHuberTavakoli2023,LiHuberFriis2025} or obtainable from full information on the density operator~\cite{HulpkeBrussLewensteinSanpera2004, Eltschka2013, Johnston2013, ZhangDenkerAsadianGuehne2024} that could be employed after local projections to finite-dimensional subspaces. The only exception is the upper bound on the SN detectable with our method given by the SN certified from the exact fidelity. 

This article is structured as follows: In Sec.~\ref{sec:framework} we present all technical prerequisites, including the notation and quantities of interest to describe CV systems and HD entanglement, in Secs.~\ref{sec:continuous variables} and~\ref{sec:schmidtnumberandhowto}, respectively. Section~\ref{sec:SN detection from CM} describes the different methods we study to detect the SN based only on the CM of the unknown state: Our new fidelity bound for SN detection is introduced in Sec.~\ref{sec:fbound}, while Sec.~\ref{sec:SN and EOF} describes the EoF, its estimation from the CM, and how it can be used to bound the SN. Section~\ref{sec:aves bound} reviews the linear Gaussian SN witnesses from~\cite{ShahandehSperlingVogel2013}. In Sec.~\ref{sec:results} we then apply these methods to different examples: Section~\ref{sec:TMST states} considers the pure Gaussian TMST states, whereas Sec.~\ref{sec:PAD TMSV states} provides a description of PAD TMSV states and lower bounds on their SN. Finally, we discuss our results and present conclusions in Sec.~\ref{sec:discussion}.


\section{Framework}\label{sec:framework}

{\noindent}Before we discuss the SN witnesses in Sec.~\ref{sec:SN detection from CM}, we now provide the relevant context and basic definitions. Readers familiar with these concepts might wish to skip ahead to Sec.~\ref{sec:fbound} and enjoy the fruits of our labor with a good glass of wine or another beverage of their choice. In Sec.~\ref{sec:continuous variables} we discuss continuous variables, Gaussian and non-Gaussian states, and establish the conventions we use throughout for mode and quadrature operators, as well as their first and second statistical moments (in particular, for the CM). Section~\ref{sec:schmidtnumberandhowto} briefly reviews the concept of HD entanglement and the SN as well as pertinent bounds on it.  


\subsection{Continuous Variables}\label{sec:continuous variables}

{\noindent}In this section we introduce the basic concepts required for the description of CV systems in this work. For more detailed reviews on the subject we point to~\cite{BraunsteinVanLoock2005, WeedbrookPirandolaGarciaPatronCerfRalphShapiroLloyd2012, AdessoRagyLee2014} or~\cite[Chapter~25]{BertlmannFriis2023}.

A CV system consists of $M$ subsystems that are referred to as modes, each of which is associated with the degrees of freedom of a quantum-mechanical harmonic oscillator. 
For each mode, labeled by an index $i=1,2,\ldots,M$, we can define the creation and annihilation operators $\hat{a}^\dagger_i$ and $\hat{a}^{\phantom{\dagger}}_{i}$, respectively. 
From these, one constructs the quadrature operators $\hat{x}_{i}$ and $\hat{p}_{i}$:
\begin{subequations}\label{eq:quadrature operators}
\begin{align}
    \hat{x}_{i}&=  \tfrac{1}{ \sqrt{2}}(\hat{a}^{\phantom{\dagger}}_{i}+ \hat{a}^{\dagger}_{i}),
    \label{eq:quadrature pos}
    \\ 
     \hat{p}_{i}&=  \tfrac{-\mathrm{i}}{ \sqrt{2}}(\hat{a}^{\phantom{\dagger}}_{i}-\hat{a}^{\dagger}_{i}),
     \label{eq:quadrature momentum}
\end{align}
\end{subequations}
which satisfy the canonical commutation relations
\begin{align}
    [\hat{x}_{i},\hat{p}_{j}]= \mathrm{i}  \delta_{ij} .
    \label{eq:commuatationrelation}
\end{align}
The operators can be arranged into a vector $\hat{r}$ with $2M$ components, 
\begin{align}
    \hat{r}&
    =\,(\hat{x}_{1},\hat{p}_{1},\hat{x}_{2},\hat{p}_{2},\ldots,\hat{x}_{M},\hat{p}_{M})^{\intercal}.
\end{align}
The commutation relations for all modes can now be compactly expressed as $[\hat{r}_{k},\hat{r}_{l}]= \mathrm{i}\, \Omega_{kl}$ for $k,l=1,2,\ldots,2M$,  
where $\Omega_{kl}$ are the components of the symplectic form $\Omega$ defined by
\begin{align}
      \Omega=  \bigoplus_{i=1}^{M}
       \begin{pmatrix}
           0&1\\
           -1&0
       \end{pmatrix}.
\end{align}

Any CV state~$\rho$ can be expressed in terms of its Wigner function. In this work, states whose Wigner function is a multivariate Gaussian are called Gaussian states. For such states, the Wigner function and thus the state itself is fully defined by the first and second statistical moments of the quadrature operators. 
These moments are given by the vector $\xi(\rho)= \langle \hat{r} \rangle_{\rho}$ and by the CM~$\gamma$, respectively. 
The CM of an $M$-mode system is a real and symmetric $2M\times2M$ matrix with components
\begin{align}
   \gamma_{kl}(\rho) &=\, \langle \hat{r}_{k}\hat{r}_{l}+\hat{r}_{l}\hat{r}_{k} \rangle_{\rho} -2 \langle \hat{r}_{k} \rangle_{\rho}  \langle \hat{r}_{l} \rangle_{\rho}  .\label{eq:cmg}
\end{align}
Here, we choose the same convention for the definition of the CM as in Ref.~\cite{BaksovaLeskovjanovaMistaAgudeloFriis2025}.
In other references, in particular, in Ref.~\cite{ShahandehSperlingVogel2013} which we frequently refer to in this work, the CM is defined with an additional factor of one half. 
In our conventions, the CM satisfies the uncertainty relation $\gamma+\mathrm{i}\,\Omega\geq 0$, and, conversely, every real, symmetric matrix~$\gamma$ that satisfies this bona-fide condition is a valid CM to which a Gaussian state can be associated~\cite{SimonMukundaDutta1994}. 
Using local symplectic transformations, the CM of any pair of modes can be transformed to its standard form~\cite{DuanGiedkeCiracZoller2000}, which is given by 
\begin{align}
    \gamma &=\,
    \begin{pmatrix}
    n & 0 & c_{x} & 0\\
    0 & n & 0 & c_{p}\\
    c_{x} & 0 & m & 0\\
    0 & c_{p} & 0 & m
    \end{pmatrix},
    \label{standardform}
\end{align}
where, $n\geq m\ge1$ and $c_{x}\geq -c_{p}>0$. Since local symplectic transformations can be realized by local (Gaussian) unitaries~\cite{serafini2023, BertlmannFriis2023} on the underlying state, they do not change the amount of entanglement and leave the Gaussian character of the state unchanged. Consequently, the standard-form CM fully describes the entanglement of any two-mode Gaussian state. 

Here, we call every state non-Gaussian whose Wigner function is not Gaussian, as opposed to stronger conditions for non-Gaussianity such as the inability to decompose the Wigner function into a convex sum of Gaussian Wigner functions~\cite{FilipMista2011} or a Wigner function that has negative values~\cite{KenfackZyczkowski2004}. Irrespective of this convention, the higher statistical moments of any non-Gaussian state encode information about the state and its correlations that is independent of the CM.

For any $M$-mode state~$\rho$, the mean excitation number~$N$ is given by 
\begin{align}
    N(\rho) &=\,\sum\limits_{i=1}^{M}\expval{\hat{a}_{i}^{\dagger}\hat{a}^{\phantom{\dagger}}_{i}\!}_{\rho}
    \,=\,
    \tfrac{1}{4}\tr\bigl[\gamma(\rho)\bigr]\,-\,\tfrac{M}{2}
    \,+\,\tfrac{1}{2}\|\xi(\rho)\|^{2}.
    \label{eq:meanphoton}
\end{align}
The mean excitation number has independent contributions from the CM $\gamma(\rho)$ and the vector of first moments $\xi(\rho)$, but it is worth keeping in mind that the first moments can be adjusted at will by local unitary displacements, which do not change the CM or the entanglement of the state.


\subsection{The Schmidt Number and How to Bound It}\label{sec:schmidtnumberandhowto}

{\noindent}Entanglement is widely considered to be a key resource enabling quantum advantages in quantum communication tasks~\cite{Ekert1991, BennettBrassardPopescuSchumacherSmolinWooters1996, JozsaLinden2003, GiovannettiLloydMaccone2006}. However, the mere presence of entanglement alone is not sufficient for a task at hand; both the quality and quantity of the entanglement present in a quantum system play a crucial role for its usefulness. An important quantifier of bipartite entanglement is the SN, sometimes referred to as the entanglement dimension~\cite{TerhalHorodecki2000}. For a bipartite pure state $\ket{\psi}\in\mathcal{H}_{A}\otimes\mathcal{H}_{B}$, the Schmidt rank (SR) is given by the rank of the reduced states, $\mathrm{SR}(\psi)=\operatorname{rank}\bigl(\tr_{B}(\ket{\psi}\!\!\bra{\psi})\bigr)=\operatorname{rank}\bigl(\tr_{A}(\ket{\psi}\!\!\bra{\psi})\bigr)$, and describes the number of entangled degrees of freedom of a system, i.e., the number of non-zero terms in the Schmidt decomposition of $\ket{\psi}$. The SN of a mixed state~$\rho$ is the highest SR appearing in its decomposition into pure states, minimized over all such decompositions, 
\begin{align}
    \operatorname{SN}(\rho):= \min_{\mathcal{D}(\rho)}  \max_i \Big\{\text{SR}(\psi_i):\ \rho=\sum_i p_i |\psi_i\rangle\!\langle\psi_i|\Big\},
    \label{eq:Schmidt}
\end{align}
where $\mathcal{D}(\rho)=\{(p_i,\ket{\psi}_i):\rho=\sum_i p_i |\psi_i\rangle\!\langle\psi_i|\}$ is the set of all decompositions of~$\rho$, with $0\leq p_i\leq 1$ and $\sum_i p_i=1$.\\[-2.5mm]

The set of all states with $\operatorname{SN}(\rho)\le{k}$ forms a convex set, leading to a nested structure of the state space, where the innermost set consists of all separable states with SN$(\rho)=1$. 
This convexity means that the detection of SNs $k>2$ follows similar principles as the detection of the presence of entanglement (SN $k>1$) itself. 
It is therefore perhaps not surprising that known techniques for entanglement detection can be adapted to the detection of arbitrary SNs. We now review some pertinent examples.\\[-2.5mm]

To understand how the fidelity of a CV state with respect to a reference can be used to bound the SN, let us consider the discrete-variable case first. For bipartite systems with subsystems $A$ and $B$ of finite and equal local dimension $d$, a family of SN witnesses can be obtained from evaluating correlations in sets of $\mathcal{M}\geq2$ local orthonormal bases $\{\ket{\psi\suptiny{0}{0}{(j)}_{i}}_{A}\}_{i=0,\ldots,d-1}$ and $\{\ket{\psi\suptiny{0}{0}{(j)}_{i}}_{B}\}_{i=0,\ldots,d-1}$ for the Hilbert spaces of $A$ and $B$, respectively, where $j=1,\ldots,\mathcal{M}$ labels the different bases. 
For such a chosen set of bases and a given bipartite state~$\rho$, we define the quantity
\begin{align}
    \mathcal{S}_{d}\suptiny{1}{0}{(\mathcal{M})}(\rho)
    &\coloneq\,
    \sum\limits_{j=1}^{\mathcal{M}}\sum\limits_{i=0}^{d-1}
 \bra{\psi\suptiny{0}{0}{(j)}_{i},{\psi\suptiny{0}{0}{(j)}_{i}}^{*}}\rho\ket{\psi\suptiny{0}{0}{(j)}_{i},{\psi\suptiny{0}{0}{(j)}_{i}}^{*}}.
\end{align}
Now, for any state~$\rho$ with SN at most~$k$ the quantity $\mathcal{S}_{d}\suptiny{1}{0}{(\mathcal{M})}(\rho)$ satisfies the condition~\cite{LiHuberFriis2025}
\begin{align}
    \mathcal{S}_{d}\suptiny{1}{0}{(\mathcal{M})}(\rho) &\leq\, \tfrac{k\,[\mathcal{M}-\tilde{\mathcal{M}}(\mathcal{C})]}{d}\,+\, \tilde{\mathcal{M}}(\mathcal{C}),
    \label{eq:LiHubFri witness}
\end{align}
where $\tilde{\mathcal{M}}(\mathcal{C}) := \min\{\lambda(\mathcal{C}), \mathcal{M}\}
$, with
\begin{align}
    \lambda(\mathcal{C}) &\coloneq\, \tfrac{1}{2}\bigl(1+
    \bigr[
    1+2d\sum_{j\neq j'}G\suptiny{1}{0}{(j,j')}
    ]^{1/2}
    \bigr)\,\geq\, 1\,,
\end{align}
which only depends on the overlaps of the different basis states, collected in the set 
\begin{align}
    \mathcal{C} &\coloneq\,\{|\!\scpr{\psi\suptiny{0}{0}{(j)}_{i}}{\psi\suptiny{0}{0}{(j')}_{i'}}\!|^{2}\}_{i,i';j\neq j'},
\end{align}
 via the functions
\begin{subequations}
\begin{align}
    G\suptiny{1}{0}{(j,j')} &\coloneq\, 1-(d+1)c\suptiny{0}{0}{(j,j')}_\text{min} 
    + \tfrac{1}{d}\sum_{i,i'}|\!\scpr{\psi\suptiny{0}{0}{(j)}_{i}}{\psi\suptiny{0}{0}{(j')}_{i'}}\!|^{4},\\
    c\suptiny{0}{0}{(j,j')}_\text{min}  &\coloneq\,\min_{i,i'} |\!\scpr{\psi\suptiny{0}{0}{(j)}_{i}}{\psi\suptiny{0}{0}{(j')}_{i'}}\!|^{2}\,.
\end{align}
\end{subequations}
Any state that violates the inequality~(\ref{eq:LiHubFri witness}) must hence have a SN of at least~$k+1$.\\[-2.5mm] 

In the case that the local orthonormal bases are mutually unbiased, i.e., if they satisfy 
\begin{align}
    |\!\scpr{\psi\suptiny{0}{0}{(j)}_{i}}{\psi\suptiny{0}{0}{(j')}_{i'}}\!|^{2}
    &=\,\tfrac{1}{d}
\end{align} 
for all~$i$ and~$i'$ and all $j\neq j'$, one has  $\tilde{\mathcal{M}}(\mathcal{C})=1$ and the condition in~(\ref{eq:LiHubFri witness}) reduces to the SN witness from~\cite{MorelliHuberTavakoli2023}, which in turn matches the condition from~\cite{SpenglerHuberBrierleyAdaktylosHiesmayr2012} when testing for the presence of entanglement (i.e., for $k=1$).\\[-2.5mm] 

The methods described above can be applied for SN detection if measurements in arbitrary bases are available, which provides some flexibility to adjust to practical restrictions, but generally work best for MUBs~\cite{LiHuberFriis2025}. In addition, the maximal fidelity $\mathcal{F}_{\mathrm{max}}(\rho)$ of the state~$\rho$ with any maximally entangled state, given by
\begin{align}
    \mathcal{F}_{\mathrm{max}}(\rho) &\coloneq\,  \max_{U} \bra{\phi^+_d}U\otimes\mathbb{1}\ \rho\ U^\dagger \otimes \mathbb{1}\ket{\phi^+_d}\,,
\end{align}
where $|\phi^+_d\rangle=(1/\sqrt{d})\sum_{i=0}^{d-1}|i,i\rangle$ is a maximally entangled state in local dimension~$d$ and $U$ are (local) unitaries, can be bounded from below by~\cite{LiHuberFriis2025}
\begin{align}
    \mathcal{F}_{\mathrm{max}}(\rho) &
    \geq\,
    \max\Bigl\{
    0,\tfrac{\mathcal{S}_{d}\suptiny{1}{0}{(\mathcal{M})}(\rho)-\tilde{\mathcal{M}}(\mathcal{C})}{\mathcal{M}-\tilde{\mathcal{M}}(\mathcal{C})}
    \Bigr\}\,.
\end{align}
A value of $\mathcal{F}_{\mathrm{max}}(\rho)$ larger than $k/d$ then implies a SN of~$\rho$ of $\operatorname{SN}(\rho)\geq k+1$~\cite{TerhalHorodecki2000}.\\[-2.5mm]

The fidelity with an arbitrary pure state can also be used more generally to detect the SN of a given state, which we will phrase in the following lemma:
 \begin{lemma}\label{lemma:fidelity with SN-r state}
    For any state~$\rho$ with SN~$k$, the fidelity $\mathcal{F}\bigl(\rho, \ket{\psi}\bigr)$ with any pure
    state $\ket{\psi}$ is bounded by
     \begin{align}
         \mathcal{F}\bigl(\rho, \ket{\psi}\bigr) &\leq\, 
         B_{k}\bigl(\ket{\psi}\bigr)\,\coloneq\, 
         \sum_{i=0}^{k-1} \lambda_{i}^{2},
    \label{eq:schmidtnumber}
    \end{align}
    where the summation is over the squares of the~$k$ largest Schmidt coefficients of the pure target state $\ket{\psi}$. Hence, if $ \mathcal{F}\bigl(\rho, \ket{\psi}\bigr)\,>\,B_{k}\bigl(\ket{\psi}\bigr)$, the SN of~$\rho$ is at least $k+1$.
\end{lemma}

For a proof of Lemma~\ref{lemma:fidelity with SN-r state} we refer to~\cite[Supplementary Material, Sec. C]{FicklerEtAl2014}, where this result was first reported as far as we are aware, or to~\cite[Theorem 17.3]{BertlmannFriis2023}.\\[-2.5mm] 

What Lemma~\ref{lemma:fidelity with SN-r state} leaves open is how an estimate of the desired fidelity is obtained in practice. In~\cite{BavarescoEtAl2018} it was shown that measurements in as little as two pairs of local bases can be sufficient to provide a lower bound on fidelities to pertinent target states: An adaptive strategy can be employed based on measurements in a first basis pair to construct and measure with respect to one or more additional sets of vectors, which are generally not orthonormal. Alternatively, a number of fixed MUBs can be chosen, providing lower bounds on the fidelity of~$\rho$ with $\ket{\phi^+_d}$. This approach has been successfully employed in experiments~\cite{BavarescoEtAl2018, Ecker-Huber2019,  HerreraValenciaSrivastavPivoluskaHuberFriisMcCutcheonMalik2020, ChangSarihanLiEtAl2026} and has the advantage that it also provides a lower bound on the EoF.\\[-2.5mm] 

Aside from the techniques described above, methods for SN detection in finite-dimensional systems have been developed based on symmetric sets of projectors, including complete sets of MUBs and symmetric, informationally complete projective operator-valued measures (POVMs)~\cite{MorelliHuberTavakoli2023, TavakoliMorelli2024}, later extended to more general sets of symmetric measurements~\cite{WangSunFeiWang2024, WangFei2025, Siudzinska2025} and arbitrary bases~\cite{LiHuberFriis2025}. Meanwhile, also the well-known criteria based on positivity under partial transposition~(PPT)~\cite{Peres1996,Horodecki1997} and computable cross-norm or realignment~(CCNR)~\cite{ChenWu2003, Rudolph2003, Rudolph2005} can be extended to SN criteria~\cite{HulpkeBrussLewensteinSanpera2004, Eltschka2013, Johnston2013, ZhangDenkerAsadianGuehne2024}. Similarly, tests based on CMs~\cite{GuehneHyllusGittsovichEisert2007,GittsovichGuehne2010} can be extended to SN witness~\cite{LiuHeHuberGuehneVitagliano2023,LiuFadelHeHuberVitagliano2024}, and also correlation-tensor norms can be used to detect high-dimensional entanglement~\cite{KloecklHuber2015}.\\[-2.5mm]

What unites all of these approaches is that they are difficult to employ in CV systems, either because they require full state tomography to reconstruct the density operator, or at least measurements in a number of carefully selected bases in finite-dimensional subspaces. While the former are essentially out of the question in infinite-dimensional systems, the latter can be realized in principle but may still present a challenge in practice. For instance, for a typical entangled Gaussian state like a TMSV state, measurements that resolve excitation numbers and measurements in specific superpositions of Fock states (up to some cut-off excitation number) would be required.\\[-2.5mm] 

The few examples of SN witnesses tailored specifically to CV systems are either based on covariances of the quadrature operators~\cite{ShahandehSperlingVogel2013} or on those of Bloch operators~\cite{LiuGuoFadelHeHuberVitagliano2025}. Here, we focus on the information encoded in the CM of the quadratures. We develop a new approach to estimate lower bounds on the fidelity to pure Gaussian states, and obtain related SN bounds using Lemma~\ref{lemma:fidelity with SN-r state}. We compare these to other SN bounds based on the CM, including the method of Ref.~\cite{ShahandehSperlingVogel2013}, and one based on lower bounds on the EoF. All three methods are explained in the next section.

\vspace*{-1mm}
\section{Schmidt-Number Detection from the Covariance Matrix}\label{sec:SN detection from CM}

\vspace*{-1mm}
\subsection{Fidelity Bounds and Schmidt-Number Witnesses from Quadrature Measurements}\label{sec:fbound}
\vspace*{-1mm}

{\noindent}Due to the infinite-dimensional nature of CV systems, estimating the fidelity to CV states can be experimentally costly and might require tailoring of the measurements to specific target states~\cite{MeleMeleBittelEisertGiovannettiLamiLeoneOliviero2025, FawziOufkirSalzmann2026}. 
However, as we shall show now, lower bounds on the fidelity with arbitrary pure Gaussian states can be obtained from estimates of the CM resulting from fixed measurements of the quadratures. 
To arrive at these bounds, we use the recently derived upper bounds on the trace distance: It was shown in~\cite{MeleMeleBittelEisertGiovannettiLamiLeoneOliviero2025} that the trace distance between an arbitrary state~$\rho$ and a pure Gaussian state $\ket{\psi}$, given by
\begin{align}
    \mathcal{T}(\rho,\ket{\psi}) &\coloneq\,\tfrac{1}{2}\|\rho-\ket{\psi}\!\!\bra{\psi}\|\subtiny{0}{-1}{1}\,, 
    \label{eq:tracedistance}
\end{align}
where $\|\sigma\|\subtiny{0}{0}{1}=\tr\sqrt{\sigma^{\dagger}\sigma}$ is the trace norm of an operator~$\sigma$, can be bounded from above using a quantity $\mathcal{R}(\rho,\ket{\psi})$. It depends on the first-moment vectors, $\xi(\rho)$ and $\xi(\ket{\psi})$, the CMs, $\gamma(\rho)$ and $\gamma(\ket{\psi})$, and the larger of the two average excitation numbers $N\subtiny{0}{0}{\mathrm{max}}\coloneq\,\max\{N(\rho),N(\ket{\psi})\}$:
\begin{align}
    \mathcal{R}(\rho,\ket{\psi})& \coloneq 
    \sqrt{N\subtiny{0}{0}{\mathrm{max}}\!+\! \tfrac{M}{2}} 
    \ \times \label{qvalue}\\
    &\ \ \times\sqrt{
    2\nr\|\xi(\rho)\nl-\nl \xi(\ket{\psi})\|\subtiny{0}{-1}{2}^2
    +\|\gamma(\rho)\nl -\nl \gamma(\ket{\psi})\|\subtiny{0}{-1}{\infty}
    }\,, \nonumber
\end{align}
where $\|\xi(\rho)\nl-\nl \xi(\ket{\psi})\|\subtiny{0}{-1}{2}$ is the Euclidean distance between the first moments, and $\|\sigma\|\subtiny{0}{-1}{\infty}$ is the operator norm of~$\sigma$, given by the absolute value of the largest singular value of~$\sigma$. The trace distance between the state~$\rho$ and a pure Gaussian state~$\ket{\psi}$ is then bounded by
\begin{align}
    \mathcal{T}(\rho,\ket{\psi}) &\leq\,  \min\{1, \mathcal{R}(\rho,\ket{\psi})\}\, .
    \label{eq:tracedismin}
\end{align}

We now convert this trace-distance inequality into an inequality for the fidelity by leveraging 
the Fuchs{\textendash}van~de~Graaf inequalities~\cite{FuchsVanDeGraaf1999}, which provide upper and lower bounds on the fidelity $\mathcal{F}(\rho,\ket{\psi})$ based on the trace distance $\mathcal{T}(\rho,\ket{\psi})$, i.e.,
\begin{align}
    1- \sqrt{\mathcal{F}(\rho,\ket{\psi})} \,\leq\, \mathcal{T}(\rho,\ket{\psi}) \,\leq\,  \sqrt{1-\mathcal{F}(\rho,\ket{\psi})}. 
    \label{eq:vandegraaf}
\end{align}
Both the fidelity and trace distance lie in the interval $[0,1]$. All expressions in~(\ref{eq:vandegraaf}) are thus non-negative and we can rearrange them to yield bounds on the fidelity, 
\begin{align}
    \bigl[1-\mathcal{T}(\rho,\ket{\psi})\bigr]^2 &\leq\, \mathcal{F}(\rho,\ket{\psi}) \,\leq\, 1-\mathcal{T}(\rho,\ket{\psi})^2\,.
    \label{eq:vandegraaffidelity}
\end{align}
Combining Eqs.~(\ref{eq:tracedismin}) and~(\ref{eq:vandegraaffidelity}), we thus get an inequality bounding the fidelity between an arbitrary state~$\rho$ and a pure Gaussian state $\ket{\psi}$, which is of the form 
\begin{align}
    \mathcal{F}(\rho,\ket{\psi}) &\geq\,  \bigl[1-\min\{1, \mathcal{R}(\rho,\ket{\psi})\}\bigr]^2\,, 
    \label{eq:fidelityinequality}
\end{align}
and only depends on the first and second moments of~$\rho$ and~$\ket{\psi}$.\\

We can improve any entanglement bound derived from~\eqref{eq:fidelityinequality}, e.g., via Lemma~\ref{lemma:fidelity with SN-r state}, by removing the first-moment contributions of~$\rho$ and~$\ket{\psi}$ to~$\mathcal{R}(\rho,\ket{\psi})$ and~$N\subtiny{0}{0}{\mathrm{max}}$. The corresponding operation is equivalent to local displacements, creating states $\rho\rightarrow \tilde\rho$ and $\psi\rightarrow \tilde\psi$, leaving the entanglement and CMs of the involved states invariant. 
Setting $\xi(\ket{\tilde\psi})=\xi(\tilde\rho)=0$ and using Eq.~(\ref{eq:meanphoton}), simplifies the expression $N\subtiny{0}{0}{\mathrm{max}}\!+\! \tfrac{M}{2}$, so it can be replaced by
\begin{align}
    \tilde{N}(\rho,\ket{\psi}) & \coloneq\,\tfrac{1}{4}\max\bigl\{
    \tr[\gamma(\rho)], \tr[\gamma(\ket{\psi})]
    \bigr\}\,.
    \label{eq:meanexcitationnumberwithzerofirstmoments}
\end{align}
We thus arrive at the following result: 

\begin{theorem}\label{theorem:SN detection from fid bound}
Any $M$-mode state~$\rho$ with CM $\gamma(\rho)$ that violates the inequality
\begin{align}
    \tilde{\mathcal{F}}(\rho,\ket{\psi})\,\coloneq\,\bigl[1-\min\bigl\{1, 
    \tilde{\mathcal{R}}(\rho,\ket{\psi})
    \bigr\}\bigr]^2 &\leq\,B_{k}(\ket{\psi}) 
    \label{eq:fidelityinequality refined}
\end{align}
with $B_{k}(\ket{\psi})$ given in Lemma~\ref{lemma:fidelity with SN-r state}, and 
\begin{align}
    \tilde{\mathcal{R}}(\rho,\ket{\psi}) &\coloneq\,
    \sqrt{\tilde{N}(\rho,\ket{\psi}) \,
    \|\gamma(\rho)\nl -\nl \gamma(\ket{\psi})\|\subtiny{0}{-1}{\infty}^{\phantom{2}}}
    \label{eq:Rtilde}
\end{align}
for some pure Gaussian state $\ket{\psi}$ with CM $\gamma(\ket{\psi})$ must have a SN of at least~$k+1$.
\end{theorem}

We will investigate the usefulness of the criterion in Theorem~\ref{theorem:SN detection from fid bound} 
for both pure and mixed, as well as Gaussian and non-Gaussian states in Sec.~\ref{sec:results}, and compare the results to other methods for detecting the SN based only on the CM that we will discuss in detail in Secs.~\ref{sec:SN and EOF} and~\ref{sec:aves bound}.  

\vspace*{-1mm}
\subsection{Schmidt Number and Entanglement of Formation}\label{sec:SN and EOF}
\vspace*{-1mm}

{\noindent}The entanglement of formation (EoF) of an arbitrary state~$\rho$, first introduced in~\cite{BennettDiVincenzoSmolinWootters1996}, is defined as 
\begin{align}
\operatorname{E}_{\mathrm{F}}(\rho) &=\, \inf\limits_{\mathcal{D}(\rho)}\, \sum\limits_{i} p_{i}\, \operatorname{E}(\ket{\psi_i}),
\end{align}
where $\mathcal{D}(\rho)$ is the set of all decompositions of~$\rho$ as in Eq.~(\ref{eq:Schmidt}), and $\operatorname{E}(\ket{\psi_i})$ is the entropy of entanglement of the pure states $\ket{\psi_i}$, given by the von Neumann entropy of the reduced states, $\operatorname{E}(\ket{\psi})=S_{\mathrm{vN}}(\rho\subtiny{0}{0}{A})=S_{\mathrm{vN}}(\rho\subtiny{0}{0}{B})$ with the von~Neumann entropy $S_{\mathrm{vN}}(\rho)=-\tr[\rho\log(\rho)]$ for $\rho\subtiny{0}{0}{A/B}=\tr\subtiny{0}{0}{B/A}(\ket{\psi}\!\!\bra{\psi})$. 
The EoF is a proper entanglement measure in the sense that it vanishes for all separable states, is normalizable, is non-increasing on average under local operations and classical communication (LOCC)~\cite{BennettDiVincenzoSmolinWootters1996}, and reduces to the entropy of entanglement for pure states (see, e.g.,~\cite{PlenioVirmani2007} or~\cite[Chapter~16]{BertlmannFriis2023} for reviews on entanglement measures). The optimization over all decompositions is generally hard to solve and analytical solutions are only known for a few special cases, including arbitrary states of two qubits~\cite{Wootters1998}, states invariant under certain groups of local symmetries in arbitrary finite dimensions~\cite{TerhalVollbrecht2000, VollbrechtWerner2001}, and arbitrary two-mode Gaussian states in CV systems~\cite{AkbariKourbolaghAlijanzadehBoura2015}.\\ 

Here, we are interested in the EoF primarily because it provides a lower bound on the SN of a given state~$\rho$ via the inequality
\begin{align}
    \log \bigl[\operatorname{SN}(\rho)\bigr] &\geq\, \operatorname{E}_{\mathrm{F}}(\rho) .
    \label{eq:SNfromEoF}
\end{align}
To see that this is the case, we note that for every bipartite state~$\rho$ with SN~$k$ there exists a decomposition $\rho=\sum_{i}p_{i}\ket{\psi_{i}}\!\!\bra{\psi_{i}}$ such that $\operatorname{SN}(\rho)=k=\max_{i}\operatorname{SR}(\ket{\psi_{i}})$. Now, for every $\ket{\psi_{i}}$ in this decomposition with $\operatorname{SR}(\ket{\psi_{i}})=k_i\leq k$, the entropy of entanglement, given by the subsystem entropy for the pure states considered here, is bounded by
\begin{align}
    \operatorname{E}(\ket{\psi_{i}}) & 
    \leq\, \log(k_{i}) \,\leq\, \log(k)\,=\,\log
    \bigl[\operatorname{SN}(\rho)\bigr].
    \label{eq:krelation}
\end{align}
The EoF of~$\rho$, on the other hand, includes a minimization over all decompositions of~$\rho$, and so is smaller or equal than the average entropy of entanglement across the ensemble that attains the infimum for the SN, and therefore 
\begin{align}
   \operatorname{E}_{\mathrm{F}}(\rho) &\leq 
   \sum\limits_{i}p_{i}\,\operatorname{E}(\ket{\psi_{i}})
   \,\leq\,
   \underset{i} {\mathrm{max}} \    \operatorname{E}(\ket{\psi_{i}}).
    \label{eq:EoFrelation}
\end{align}
Since the inequality in~(\ref{eq:krelation}) holds for all $\ket{\psi_{i}}$, including the one for which $k_i=k$, we can combine inequalities~(\ref{eq:krelation}) and~(\ref{eq:EoFrelation}) to obtain the expression in~(\ref{eq:SNfromEoF}).

{\noindent}The EoF thus provides a lower bound on the (logarithm of the) SN, and this lower bound can be tight, e.g., for maximally entangled states in finite dimension $d$, for which $\operatorname{E}_{\mathrm{F}}(|\phi^+_d\rangle)=\log[\operatorname{SN}_{\mathrm{F}}(|\phi^+_d\rangle)]=\log(d)$. However, there is no converse lower bound: The SN of a (pure) state can be arbitrarily large (up to the dimension $d$) while the EoF is arbitrarily close to zero. For example, consider the TMSV state given by
\begin{align}
     |\mathrm{TMSV}(s)\rangle &= \tfrac{1}{\cosh (s)}\sum^{\infty}_{n=0}\tanh^{n}(s)\ket{n}_1\ket{n}_2,
      \label{eq:tmsv}
\end{align}
where $s$ is the squeezing parameter of the state and $\ket{n}_i$ is a Fock state obtained by $n$-fold application of the creation operator $a^{\dagger}_i$ to the vacuum $\ket{0}_i$ of mode $i=\{1,\,2\}$, that is, $\ket{n}_i=(1/\sqrt{n!})(a^{\dagger}_i)^{n}\ket{0}_i$. These Fock states $\ket{n}_i$ are the eigenstates of $\hat{a}_{i}^{\dagger}\hat{a}^{\phantom{\dagger}}_{i}$ such that $\hat{a}_{i}^{\dagger}\hat{a}^{\phantom{\dagger}}_{i}\ket{n}_i=n\ket{n}_i$. For any $s\neq0$ the SN of the TMSV state is infinite, while it is zero for $s=0$. In contrast, the EoF behaves smoothly, monotonically growing from zero at $s=0$ as $|s|$ increases.\\
 
Let us now focus on the specific case of the EoF for Gaussian states. For symmetric two-mode Gaussian states $\rho\subtiny{0.5}{0}{\mathrm{SG}}$, i.e., two-mode Gaussian states for which the subsystem entropies of the two modes match, a closed-form expression for the EoF was provided in~\cite{GiedkeWolfKruegerWernerCirac2003}. These states can be characterized via the standard-form CM given in Eq.~(\ref{standardform}) with $n=m$ and $c_{x}\ge-c_{p}>0$. 
The EoF is then given by 
\begin{align}\label{eq:eof}
    \operatorname{E}_{\mathrm{F}}(\rho\subtiny{0.5}{0}{\mathrm{SG}}) &=\,h(\Delta),
\end{align}
where $\Delta = \sqrt{(n-c_{x})(n+c_{p})}$ and
\begin{align}
 h(z) &=\, \tfrac{(1+z)^2}{4z}\,\log\bigl(\tfrac{(1+z)^2}{4z}\bigr)
       -\tfrac{(1-z)^2}{4z}\,\log\bigl(\tfrac{(1-z)^2}{4z}\bigr).
       \label{eq:EoFdelta}
\end{align}
This result has been extended to general (non-symmetric) two-mode Gaussian states~$\rho\subtiny{0.5}{0}{\mathrm{G}}$ where $n\neq m$: In Ref.~\cite{TserkisRalph2017} a lower bound on the EoF for general two-mode Gaussian states was provided, while Ref.~\cite{AkbariKourbolaghAlijanzadehBoura2015} presents an expression for the EoF for all two-mode Gaussian states. To evaluate the latter, the standard-form CM given in Eq.~(\ref{standardform}) is subjected to single-mode squeezing operations parameterized by $r_{1},r_{2}\geq0$, such that $\gamma\mapsto \tilde{\gamma}=S(r_1)\oplus S(r_2) \gamma S(r_1)\oplus S(r_2)$ with $S(r_i)=\diag\{r_{i}^{1/2},r_{i}^{-1/2}\}$ for $i=1,2$. These local symplectic operations leave the entanglement unchanged, but bring the standard-form CM to the form
\begin{align}
\tilde{\gamma} &=\,
\begin{pmatrix}
n r_1 & 0 & c_{x}\sqrt{r_1 r_2} & 0\\
0 & n/r_1 & 0 & c_{p}/\sqrt{r_1 r_2}\\
c_{x}\sqrt{r_1 r_2} & 0 & m r_2 & 0\\
0 & c_{p}/\sqrt{r_1 r_2} & 0 & m/r_2
\end{pmatrix}.
\end{align}
For any given set of values $n$, $m$, $c_x$, and $c_p$ one may then find pairs of values $r_1, r_2>0$ such that the 
conditions
\begin{subequations}
    \begin{align}
    (nr_1-x)(mr_2-x)    &=\,(c_{x}\sqrt{r_1r_2}-y)^2 ,
    \label{eq:nonlinear1}\\
    (n/r_1-x)(m/r_2-x) &=\,(c_{p}/\sqrt{r_1r_2}+y)^2  
    \label{eq:nonlinear2}
\end{align}
\end{subequations}
are satisfied, where $y^{2}=x^{2}-1$ and 
\begin{align}
    x   &=\,\frac{m n \left(r_2^2-r_1^2\right)}{m r_1 \left(r_2^2-1\right)+n r_2\left(1-r_1^2\right)}\,.\label{eq:nonlinear3}
\end{align}
Out of the solutions for $r_1$ and $r_2$, one selects those values which minimize $x$, and uses them to define the quantities $a_0=\sqrt{(m r_2-x)/(n r_1-x)}$ and $c_0=a_0+(1/a_0)$, along with
\begin{subequations}
\begin{align}
\Delta_0 &= 
    \tfrac{a_0 n}{2}  
    \bigl(r_1+\tfrac{1}{r_1}\bigr)
    +\tfrac{m}{2 a_0}\bigl(r_2+\tfrac{1}{r_2}\bigr) 
    -\bigl(c_{x} \sqrt{r_1 r_2}-\tfrac{c_{p}}{\sqrt{r_1 r_2}}\bigr)
    , \\[-2mm]
\Delta'_0 &=\, 
\Bigl(\Delta_0+\sqrt{\Delta_0^2-c_0^2+4 
}\,\Bigr) 
/(1+\tfrac{2}{c_0}).
\end{align}
\end{subequations}
The EoF of an arbitrary two-mode Gaussian state $\rho\subtiny{0.5}{0}{\mathrm{G}}$ is then given by
\begin{align}
    \operatorname{E}_{\mathrm{F}}(\rho\subtiny{0.5}{0}{\mathrm{G}}) = h\left(\min(\Delta'_0,1)\right).
    \label{eq:eofnonsymmetric}
\end{align}
Since the EoF is strongly super-additive, Gaussian states have the smallest EoF among all states with the same CM~\cite{WolfGiedkeCirac2006}. Consequently, the EoF of any (potentially non-Gaussian) state~$\rho$ with given CM~$\gamma$ is larger or equal than the EoF of a Gaussian state $\rho\subtiny{0}{0}{\mathrm{G}}$ with the same CM, i.e.,
\begin{align}
\operatorname{E}_{\mathrm{F}} [\rho(\gamma)] &\geq\, \operatorname{E}_{\mathrm{F}} [\rho\subtiny{0.5}{0}{\mathrm{G}}(\gamma)].
\label{EoFinequalities}
\end{align}
In combination with the SN bound of~(\ref{eq:SNfromEoF}) we thus see that the SN of any state~$\rho$ with CM~$\gamma$ is bounded by
\begin{align}
\log \bigl(\operatorname{SN}[\rho(\gamma)]\bigr) &\geq\,
\operatorname{E}_{\mathrm{F}} [\rho\subtiny{0.5}{0}{\mathrm{G}}(\gamma)]\,.
\end{align}

\vspace*{-1mm}
\subsection{Linear Gaussian Schmidt-Number Witnesses}
\label{sec:aves bound}
\vspace*{-1mm}

{\noindent}Another way to obtain a SN witness for CV states based only on the first and second statistical moments of the quadrature operators was introduced in~\cite{ShahandehSperlingVogel2013}. There, the considered witness is linear in the sense that it involves evaluating the expectation value $\langle\hat L\rangle_\rho$ of a Hermitian test operator~$\hat{L}$, which is a linear function of the density operator~$\rho$. The test operator~$\hat{L}$ is further restricted to be Gaussian, i.e., it is a polynomial of at most second order in the mode operators~$\hat{a}^{\phantom{\dagger}}_{i}$ and~$\hat{a}^\dagger_i$. The expectation value of the test operator can hence be determined from measurements of the first and second moments alone.

For a CM in the standard form of Eq.~(\ref{standardform}), the expectation value of the test operator can be expressed directly in terms of the local variances~$n$ and~$m$, and the covariances $c_{x}$ and $c_{p}$ as
\begin{align}
    \langle\hat{L}
    \rangle
    &=\,
    2\,(n\nr\nu_1+m\nr\nu_2)
    +\nu_{\mathrm{c}
    }\bigl(|c_x|+|c_p|\bigr),
\end{align}
where the parameters $\nu_1$, $\nu_2$, and $\nu_{\mathrm{c}}$ characterize the remaining freedom to choose the Gaussian test operator and can be optimized for the CM of the state under consideration. The SN thresholds associated with a given test operator are obtained by minimizing its expectation value over all pure states with SR not exceeding~$k$,
\begin{align}
    g_k &\ =\,\inf_{\mathrm{SR}(\ket{\psi})\leq k}
    \langle\psi|\hat{L}|\psi\rangle.
    \label{eq:gks}
\end{align}
The quantities $g_k$ are obtained by solving the  corresponding SN-eigenvalue equations given in~\cite{SperlingVogel2011}. Any state satisfying\vspace*{-1.5mm}
\begin{align}
    \langle\hat{L}\rangle_\rho  &<\, g_k
    \label{LineargaussianSN bound}
\end{align}
cannot be represented as a mixture of pure states with Schmidt rank at most~$k$, and consequently
\begin{align}
    \mathrm{SN}(\rho)   &\geq\, k+1.
\end{align}
The certification via these bounds only requires information on second moments in terms of the CM rather than full state tomography. Therefore, this approach provides a natural benchmark for comparison with the fidelity-based certification method and the EoF-based SN bound presented in Secs.~\ref{sec:fbound} and~\ref{sec:SN and EOF}, respectively. 

\vspace*{-1mm}
\section{Results}\label{sec:results}
\vspace*{-1mm}

{\noindent}In this section, we apply our method for SN detection based on lower bounds on the fidelity to pure Gaussian states using only quadrature measurements (see Sec.~\ref{sec:fbound}) to pertinent examples and compare its performance to the SN bounds based on the EoF and linear Gaussian SN witnesses from Secs.~\ref{sec:fbound} and~\ref{sec:SN and EOF}, respectively. As a first example, Sec.~\ref{sec:TMST states} considers TMST states, which form an important family of mixed Gaussian states. Certifying the SN of such states using only quadrature measurements is especially desirable from an experimental perspective, as these quantities are readily accessible without requiring full state tomography. We subsequently extend our analysis to a family of experimentally relevant non-Gaussian states in Sec.~\ref{sec:PAD TMSV states}, thereby illustrating the applicability of our certification method beyond the Gaussian regime. 


\begin{figure*}[t!]
   \centering 
   \includegraphics[width=1\linewidth,trim={0.4cm 0cm 0cm 0cm},clip]{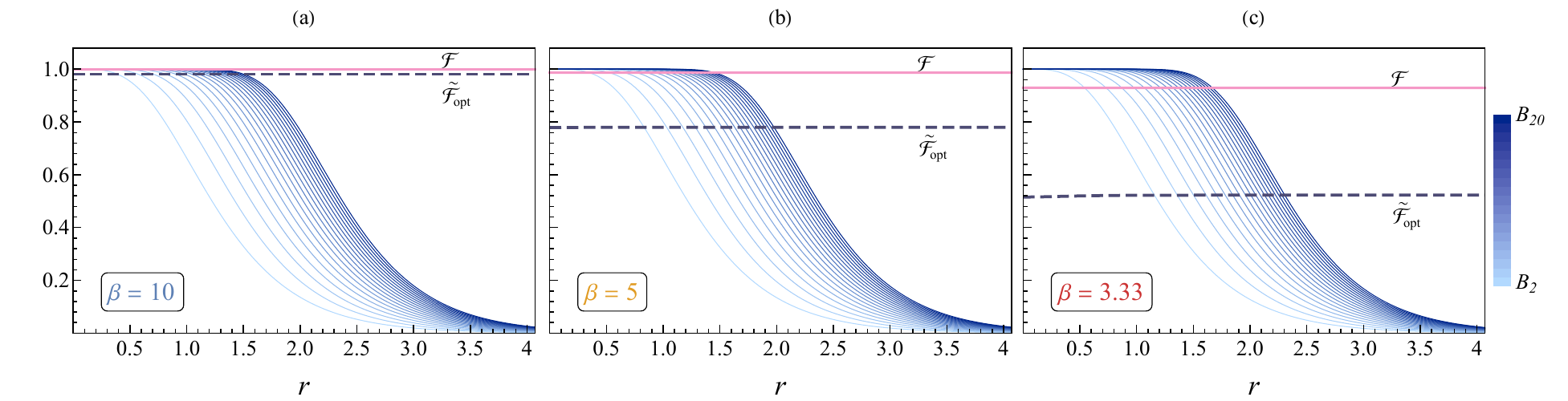}
   \vspace*{-6mm}
    \caption{\textbf{Fidelities and SN thresholds for two-mode squeezed thermal states}. The exact fidelity $\mathcal{F}$ (pink, solid) between TMST states $\rho\subtiny{0}{0}{\mathrm{TMST}}(r,\beta)$ and the optimal target state $\ket{\mathrm{TMSV}(s=s_{\mathrm{opt}})}$ from Eq.~(\ref{eq:fidelity TMSthermal and TMSV}), the lower bound $\tilde{\mathcal{F}}_{\mathrm{opt}}$ (black, dashed) from Eq.~(\ref{eq:fmax}), and the corresponding SN thresholds $B_k$ (blue) from Eq.~(\ref{eq:mainbound}) are shown as functions of~$r$ for different values of~$\beta$ with initial-state mixedness increasing left to right:  
    (a) $\beta=10$, (b) $\beta=5$, and (c) $\beta=3.33$ 
    in units of $k\subtiny{0}{0}{\mathrm{B}}/(\hbar\omega)$. 
    Although both~$\mathcal{F}$ and~$\tilde{\mathcal{F}}_{\mathrm{opt}}$ depend on~$\beta$ and~$r$, the former only via $s_{\mathrm{opt}}=s_{\mathrm{opt}}(r,\beta)$, the changes with varying~$r$ but fixed~$\beta$ are difficult to resolve for the high-purity regime we are interested in here. For instance, the relative change of~$\mathcal{F}$ across the parameter range in panel (c) is $\bigl(\mathcal{F}[s_{\mathrm{opt}}(r=0,\beta=3.33)]-\mathcal{F}[s_{\mathrm{opt}}(r=4,\beta=3.33)]\bigr)/\mathcal{F}[s_{\mathrm{opt}}(r=4,\beta=3.33)]$ is $0.128\%$.
    }
\label{fig:fig1}
\end{figure*}

\vspace*{-1.5mm}
\subsection{Two-Mode Squeezed Thermal States}\label{sec:TMST states}
\vspace*{-1.5mm}

{\noindent}The thermal state of a single mode labeled by~$i$ is given by\vspace*{-2.5mm}
\begin{align}
    \tau_i(\beta_i) =(1-e^{-\beta_i})\sum\limits_{n=0}^{\infty}e^{-\beta_i n}\ket{n}\!\!\bra{n}_i,
    \label{eq:singlemodethermal}
\end{align}
where $\beta_i$ is the inverse temperature of mode~$i$  in units of $k\subtiny{0}{0}{\mathrm{B}}/(\hbar\omega_i)$, $\omega_i$ is the (angular) frequency of the mode, and the $\ket{n}_i$ are the previously mentioned Fock states. The thermal state $\rho_{\tau}(\beta)$ of two non-interacting modes labeled by $i=1,2$ at the same inverse temperature\footnote{In our conventions, this only implies the same temperature if the two modes have the same (angular) frequencies, $\omega_1=\omega_2=\omega$.} $\beta_1=\beta_2\equiv \beta$, is given by
\begin{align}
   \rho_{\tau}(\beta) &=\,\tau_1(\beta)\otimes\tau_2(\beta). 
   \label{eq:twomodethermal}
\end{align}
Subjecting the two modes in the thermal state $\rho_{\tau}(\beta)$ to a two-mode squeezing transformation represented by the unitary operator
\begin{align}
    S\subtiny{0}{0}{\mathrm{TMS}}(r)    &=\,\exp\bigl[r(a_{1}^{\dagger}a_{2}^{\dagger}-a_{1}^{\phantom{\dagger}}a_{2}^{\phantom{\dagger}})\bigr],
\label{eq:squeezingoperator}
\end{align}
where~$r$ is the squeezing parameter, we obtain a TMST state
\begin{align}
    \rho\subtiny{0}{0}{\mathrm{TMST}}(r,\beta) &=\, S\subtiny{0}{0}{\mathrm{TMS}} (r)
    \,\rho_{\tau}(\beta)\,
    S\subtiny{0}{0}{\mathrm{TMS}}^{\dagger}(r).
\label{eq:squeezedthermalstate}
\end{align}
This TMST state will serve as the first test case for applying the previously discussed SN witnesses.\\[-2mm]

To evaluate the SN-detection method based on lower bounds to the fidelity from Sec.~\ref{sec:fbound}, we require a pure Gaussian target state $\ket{\psi}$, which we here choose to be a TMSV state as given in Eq.~(\ref{eq:tmsv}). The motivation for this choice comes naturally as for a fixed squeezing parameter $s$, TMSV states provide the entangled resource for two-mode CV systems, with their entanglement increasing monotonically with increasing magnitude of the squeezing parameter. For later comparison with the lower bound, the exact fidelity between a TMSV state and a TMST state is given by 
\begin{align}
      \mathcal{F}\bigl( 
      \rho\subtiny{0}{0}{\mathrm{TMST}}(r,\beta),
      \ket{\mathrm{TMSV}(s)} \bigr) &=\, 
      \frac{(e^{\beta}-1)^2 \ \mathrm{sech}^{\nl 2}\nl(s-r)}{e^{2\beta}-\tanh^{\nl 2}\nl(s-r)},
      \label{eq:fidelity TMSthermal and TMSV}
\end{align}
as we calculate explicitly in Appendix~\ref{appendix sec: fidelity TMSV and TMST}.\\[-2mm]

We now proceed with the analysis and optimization of the lower bound on the fidelity of Eq.~(\ref{eq:fidelityinequality refined}), with the aim of employing it for SN certification. A crucial step is to identify the optimal target state $\lvert \mathrm{TMSV}(s_{\mathrm{opt}})\rangle$ within the selected family of TMSV states for each test state with fixed parameters~$r$ and $\beta$. The target states $\ket{\mathrm{TMSV}(s)}$ are parametrized only by the squeezing~$s$ and even though the SN is always infinite for such states, an arbitrary choice of the squeezing parameter $s$ does not necessarily yield the most effective target state for certification. Indeed, the choice of the target state directly affects the tightness of the lower bound on the fidelity and, consequently, the certifiable SN. In the following, we provide analytical expressions for both the fidelity lower bound and the optimal target squeezing $s_{\mathrm{opt}}$ that maximizes the lower bound on the fidelity on the left-hand side of inequality~(\ref{eq:fidelityinequality refined}) of Theorem~\ref{theorem:SN detection from fid bound} for a given TMST state. We then use these results to certify the SN for a selected range of parameters~$r$ and~$\beta$ of the TMST states.\\[-2mm] 

The optimal squeezing parameter depends explicitly on the parameters of the TMST state and is given by
\begin{align}
    s_{\mathrm{opt}}& 
    =\,\tfrac{1}{2}\operatorname{arsinh}\bigl[\sinh(2r)\nr\coth\bigl(\tfrac{\beta}{2}\bigr)\bigr],
    \label{eq:ropt}
    \end{align}
as is derived in detail in Appendix~\ref{appendix sec: analytical solutions for TMSV ans TMST}. 
Using the optimal target-state squeezing, the optimized fidelity lower bound is then 
\begin{align}
    \tilde{\mathcal{F}}_{\mathrm{opt}}(r,\beta) 
    &\coloneq\,
    \bigl[1-\min\bigl\{1, \tilde{\mathcal{R}}_{\mathrm{opt}}(r,\beta) \bigr\}\bigr]^2,
    \label{eq:fmax}
\end{align}
with 
\begin{align}
\tilde{\mathcal{R}}_{\mathrm{opt}}(r,\beta) &=\,
\sqrt{\cosh(2r)\nr\coth\bigl(\tfrac{\beta}{2}\bigr)}
\sqrt{
e^{2r}\coth\bigl(\tfrac{\beta}{2}\bigr)-e^{2s_{\mathrm{opt}}}}\,.
\end{align}
To provide a lower bound on the SN of $\rho\subtiny{0}{0}{\mathrm{TMST}}(r,\beta)$ based on Theorem~\ref{theorem:SN detection from fid bound}, the expression in Eq.~(\ref{eq:fmax}) must be compared with the sums of the~$k$ largest squared Schmidt coefficients of the target TMSV state with the optimal squeezing parameter $s_{\mathrm{opt}}$, 
\begin{align}
       B_{k}[\ket{\mathrm{TMSV}(s_{\mathrm{opt}})}]  &:= \sum_{n=0}^{k-1} \tfrac{ \tanh^{2n}(s_{\mathrm{opt}})}{ \cosh^{2}(s_{\mathrm{opt}})}\,,
    \label{eq:mainbound}
\end{align}
with values $\tilde{\mathcal{F}}_{\mathrm{opt}}>B_{k}$ indicating a SN of (at least)~$k+1$.\\[-2mm]

\begin{figure}[hb!]
   \centering 
    \includegraphics[width=0.98\linewidth]{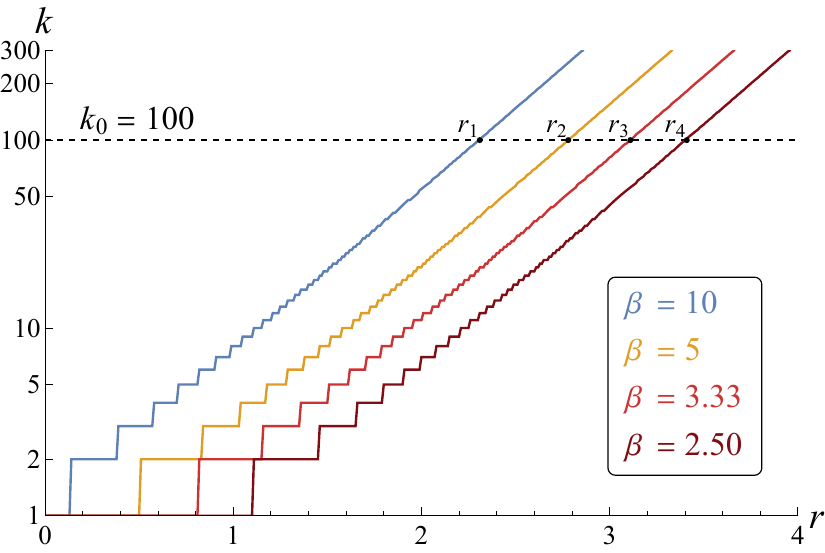}
    \vspace*{-3mm}
    \caption{\textbf{Detected Schmidt number for TMST states}: The SNs,~$k$, certified using the fidelity lower bound for TMST states in Eq.~(\ref{eq:fmax}) are shown (using a logarithmic scale on the vertical axis) as a function of~$r$ for different~$\beta$ values. The SN $k_0=100$, indicated by a dashed line, is certified at minimum squeezing $r_1=2.31$, $r_2=2.78$, $r_3=3.11$, and $r_4=3.41$  for $\beta=10$, $5$, $3.33$, and $2.5$ (left to right, top to bottom), respectively.
    }
    \label{fig: detectedSNk=100}
\end{figure}

Figures~\ref{fig:fig1} and~\ref{fig: detectedSNk=100} illustrate the behaviour of our fidelity-based SN witness for our first family of test states, the TMST states. In Fig.~\ref{fig:fig1}, we examine the lower bound $\tilde{\mathcal{F}}_{\mathrm{opt}}$ from Eq.~(\ref{eq:fmax}) as well as the exact fidelity $\mathcal{F}$ of Eq.~(\ref{eq:fidelity TMSthermal and TMSV}) between TMST states and the corresponding optimal TMSV states, relative to the SN thresholds $B_k$ from Eq.~(\ref{eq:mainbound}) as functions of~$r$ and~$\beta$. Within the chosen parameter range the thresholds $B_k$ strongly 
\begin{figure*}[ht]
   \centering 
   \includegraphics[width=1
    \linewidth]{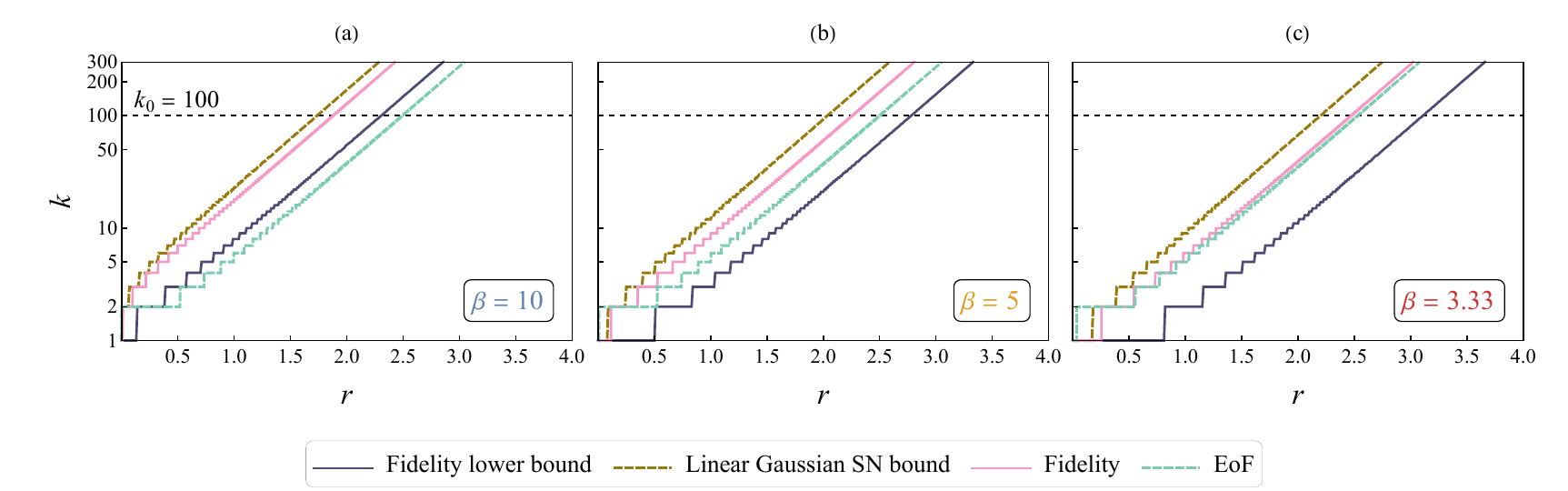} 
    \caption{\textbf{Comparisons for TMST states}: The curves show the detected SN~$k$ for TMST states based on different certification methods: linear Gaussian SN witnesses explained in Sec.~\ref{sec:aves bound} (dashed dark mustard curves), EoF-based SN witnesses as discussed in Sec.~\ref{sec:SN and EOF} (dashed mint curves), SN witnesses based on the exact fidelity (pink solid curves) between the states and corresponding optimal target state from Eq.~(\ref{eq:fidelity TMSthermal and TMSV}), and on the fidelity lower bound (solid dark-blue curves) from Eq.~(\ref{eq:fmax}). The individual panels compare the behaviour of these certification methods at different $\beta$ values: (a) $\beta=10$; (b) $\beta=5$; and (c) $\beta=0.33$. The vertical axes indicating the certified SN $k$ use a logarithmic scale, and the horizontal dashed line corresponds to the reference $k_0=100$.
    }
\label{fig3:comparionk=300allbounds}
\end{figure*}
depend on~$r$ via~$s_{\mathrm{opt}}=s_{\mathrm{opt}}(r,\beta)$ but do not significantly change with~$\beta$, while~$\mathcal{F}$ and~$\tilde{\mathcal{F}}_{\mathrm{opt}}$ do not noticeably depend on~$r$, but their values decrease and the gap between exact fidelity and lower bound widens as~$\beta$ decreases. For our first test case, we thus observe the expected behaviour of our fidelity-based SN witness: For fixed $\beta$, the detected SN increases with increasing squeezing~$r$. Conversely, to certify a fixed SN, a higher amount of squeezing is required as the thermal noise increases, as shown in Fig.~\ref{fig: detectedSNk=100}. 


\subsubsection*{Comparing Schmidt-Number Detection Methods for TMST States}\label{sec:comparisons}

{\noindent}In this section we compare different SN witnesses for TMST states to understand the connection between the limited information of the state encoded in the CM and the entanglement of the state. 

\begin{figure}[b]
   \centering \includegraphics[width=1\columnwidth]{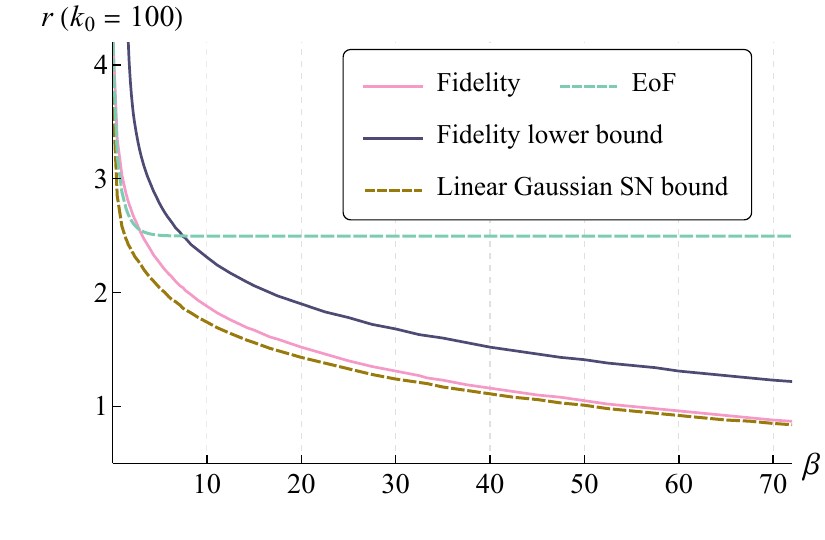}
   \vspace*{-10mm}
    \caption{\textbf{Critical squeezing for certifying $k_0$ for TMST states:} 
    The graph shows the required squeezing~$r$, needed for different bounds to detect SN $k_0=100$ for a range of $\beta$.
    } 
\label{fig4:comparionk=100}
\end{figure}

For the linear Gaussian SN witnesses discussed in Sec.~\ref{sec:aves bound}, the calculations can be simplified as explained in Appendix~\ref{appendix:LinearGaussianSNwitness}. 
As illustrated in Figs.~\ref{fig3:comparionk=300allbounds} and~\ref{fig4:comparionk=100}, for most of the parameter regions we have considered for TMST states, linear Gaussian SN witnesses provide the certification of the highest SN among all witnesses only leveraging information on the second moments of the states: That is, for TMST states characterized by the parameters~$\beta$ and squeezing~$r$, linear Gaussian SN witnesses certify larger SN than the other bounds considered, except for very small values of~$r$, where the EoF-based method sometimes performs best in detecting SN~$k=2$, see panels (b) and (c) of Fig.~\ref{fig3:comparionk=300allbounds}. Conversely, if we fix the SN of interest, for instance to $k_{0}=100$, these witnesses require the least amount of squeezing to certify a SN of~$k_{0}$ at a given~$\beta$. Nevertheless, linear Gaussian SN bounds do not provide all the information that can be accessed via the fidelity, for example closeness of the state to a pure maximally entangled target state. However, when the objective is specifically to certify SN, the fidelity-based bounds require more squeezing than linear Gaussian SN witnesses to achieve a similar level of certification. 

In this context it is important to note that, for large values of $\beta$ (small temperatures) both the exact fidelity and its lower bound are close to unity, while the SN thresholds $B_k$ lie ever closer together, the larger~$k$. Consequently, the critical squeezing~$r$ required to certify a fixed SN such as $k_{0}=100$ can differ considerably. Therefore, even a small difference between the exact fidelity and its lower bound, despite their absolute values being close, can make a substantial difference in the critical squeezing required to certify a given SN. This effect becomes particularly pronounced for the certification of large SN, where high numerical precision is required.

Finally, let us remark on the SN certification using the EoF. For TMST states, the CM is already in the standard form of Eq.~(\ref{standardform}), with $n=m=\cosh(2r)\coth\bigl(\tfrac{\beta}{2}\bigr)$ and $c_{x}=-c_{p}=\sinh(2r)\coth\bigl(\tfrac{\beta}{2}\bigr)$. To calculate the EoF, we can hence use Eq.~(\ref{eq:eof}) with $\Delta = \sqrt{(n-c_{x})(n+c_{p})}=\coth\bigl(\tfrac{\beta}{2}\bigr)e^{-2r}$. For TMST states we thus have
\begin{align}
    \operatorname{E}_{\mathrm{F}}(\rho) &=\,h(\coth(\beta/2)\,\mathrm{e}^{-2r}),
    \label{smalledsymplecticeigenvalue}
\end{align}
with the function $h$ as in Eq.~(\ref{eq:EoFdelta}).\\[-2mm]

As illustrated in Fig.~\ref{fig4:comparionk=100}, SN detection based on the EoF of Gaussian states with the same CM sometimes performs better and sometimes worse than the fidelity-based methods, but does not outperform linear Gaussian SN witnesses for TMST states. When plotting the critical squeezing required to certify a fixed SN,  $k_{0}=100$, one observes that the EoF-based method performs competitively for small inverse temperatures, $\beta$, but becomes almost constant for larger~$\beta$ in the considered range. As $\beta$ increases, the other detection methods require progressively less squeezing to be able to detect a fixed SN. For the fidelity and the fidelity lower bound this is connected to the fact that larger values of~$\beta$ imply higher purity, which in turn allows for fidelities closer to unity even for low squeezing. In this regime, the state can have a high overlap with the optimally chosen target state, making fidelity based certification effective. As the $\beta$ value decreases, the state becomes increasingly mixed and the overlap with the pure target state decreases, hence reducing the effectiveness of fidelity-based approaches. In contrast, the EoF of a state with high purity can be arbitrarily low for low squeezing, even if the state has a high SN, and is hence not particularly effective in this regime. 

Nevertheless, much like the fidelity-based approaches, the EoF-based approach has the advantage that it provides information aside from the SN, in this case a lower bound on the EoF of the state whose CM is known. 


\subsection{Photon-Added Displaced Two-Mode Squeezed Vacuum States}\label{sec:PAD TMSV states}

{\noindent}In the previous section we discussed a family of Gaussian states. Now, we focus on a family of non-Gaussian states. These states are interesting because their entanglement properties can exhibit qualitatively different behaviour from those of Gaussian states with correlations encoded in higher moments not accessible from the first and second moments alone.\\[-2mm]

Specifically, we consider a family of pure two-mode non-Gaussian states{\textemdash}PAD TMSV states{\textemdash}obtained by applying a displacement with parameter~$\alpha$ to one mode, followed by the addition of a single-photon (in the sense of the application of a creation operator and subsequent normalization) to the same mode, i.e., 
\begin{align}
  \ket{\Phi
  (\alpha,r)} &=\, \tfrac{1}{\sqrt{C}}\,\hat{a}_{1}^{\dagger}\,D_{1}\,(\alpha)\ket{\mathrm{TMSV}(r)},
    \label{eq:nongaussianstate}
\end{align}
with the normalization factor $C=1+\sinh(r)^2+\lvert \alpha \rvert^2$, where~$r$ is the squeezing parameter of the TMSV state. The displacement operator for mode~$i$ is given by
\begin{align}
    D_i(\alpha)=\exp{(\alpha \,\hat{a}_{i}^{\dagger}-\alpha^{*} \hat{a}^{\phantom{\dagger}}_{i})},
\label{eq:displacementoperator}
\end{align}
where~$\alpha\in\mathbb{C}$ is the displacement parameter and $\hat{a}_{i}^{\dagger}$ and $\hat{a}^{\phantom{\dagger}}_{i}$ are the creation and annihilation operators associated with the mode~$i$, respectively.\\[-2mm]

Photon addition to a mode~$i$ is described by the action of the creation operator~$\hat{a}_{i}^{\dagger}$, a non-unitary operation. Its application increases the excitation number~$n$ of any Fock state $\ket{n}_i$ of the mode in question by exactly one. But because it also adds a non-trivial pre-factor to every Fock state, $\hat{a}_{i}^{\dagger}\ket{n}_i=\sqrt{n+1}\ket{n}_i$, applying $\hat{a}_{i}^{\dagger}$ to an arbitrary state will change the relative magnitude of the amplitudes of different Fock states in a superposition. Consequently, the application of $\hat{a}_{i}^{\dagger}$ can modify both the average excitation number associated with the individual modes and the entanglement between them. \\[-2mm]

The displacement operator is a local unitary, and hence keeps the entanglement of the initial TMSV state invariant, but it does not commute with the creation operator $\hat{a}_{i}^{\dagger}$, and so the entanglement properties of the states $\ket{\Phi(\alpha,r)}$ in Eq.~(\ref{eq:nongaussianstate}) depend on both~$\alpha$ and~$r$.\\[-2mm]

The PAD TMSV states $\ket{\Phi(\alpha,r)}$ are non-Gaussian for all finite parameter values. However, using the identity $D_i^{\dagger}(\alpha) \hat{a}_{i}^{\dagger} D_i^{\phantom{\dagger}}(\alpha)=\hat{a}_{i}^{\dagger}+\alpha^{*}$, along with $\bra{m}_i\hat{a}_{i}\ket{n}_i=0$, one observes that the fidelity of $\ket{\Phi(\alpha,r)}$ with a suitably displaced TMSV state is
\begin{align}
    \big|\bra{\Phi(\alpha,r)} D_{1}(\alpha)\ket{\mathrm{TMSV}(r)}\big|^{2}
        &=\, \frac{\lvert \alpha \rvert^2}{C}.
    \label{eq:fidelityfornongaussian}
\end{align}
While this expression vanishes for $\alpha=0$, the fidelity converges to one in the limit $|\alpha|\rightarrow\infty$, indicating that $\ket{\Phi(\alpha,r)}$, and hence also its statistical moments, converge to the Gaussian state $\ket{\mathrm{TMSV}(r)}$ up to a (local unitary) displacement.\\[-2mm] 

\begin{figure}[ht!]
   \centering 
   \includegraphics[width=\linewidth,trim={1.1cm 0cm 0cm 0cm},clip]{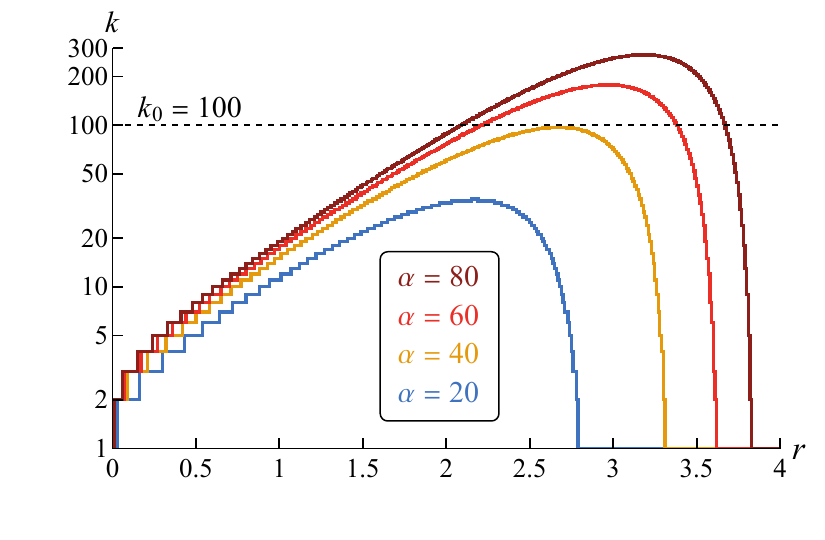}
    \vspace*{-7mm}
    \caption{\textbf{Detected Schmidt numbers for PAD TMSV states:} The plot shows the detected SN for the family of non-Gaussian states $\ket{\Phi(\alpha,r)}$ in Eq.~(\ref{eq:nongaussianstate}) as a function of the squeezing parameter~$r$ for selected values of the displacement~$\alpha=20$, $40$, $60$, and $80$ (bottom to top). The vertical axis shows the SN up to $k=300$ on a logarithmic scale. The dashed horizontal line indicates the detection of SN $k_{0}=100$. 
}
 \label{fig5:SNrvalues}
\end{figure}

As before, we are interested in a comparison of the SN-certification methods based on the second moments of the state, and, also as before, we begin with the fidelity lower bound given in Theorem~\ref{theorem:SN detection from fid bound}. For this purpose, we have to choose a suitable set of pure Gaussian target states. In the previous section, we considered the target states $\ket{\mathrm{TMSV}(s)}$, and we can do so again here. But, because the first moments of $\ket{\Phi(\alpha,r)}$, given in Appendix~\ref{appendix:nongaussianapendix}, are generally non-zero, we have to keep in mind that the fidelity bound in Eq.~(\ref{eq:fidelityinequality refined}) refers to the fidelity between $\ket{\mathrm{TMSV}(s)}$ and $D_{1}(\eta_{1})D_{2}(\eta_{2})\ket{\Phi(\alpha,r)}$, with suitable displacements $\eta_1$ and $\eta_2$ of both modes that do not change the entanglement.\\[-1mm] 

We then proceed in the following way: After evaluating the CM of the state $\ket{\Phi(\alpha,r)}$, whose components are explicitly given as functions of~$\alpha$ and~$r$ in Appendix~\ref{appendix:nongaussianapendix}, we consider the standard form of the CM as in Eq.~(\ref{standardform}), which is related to the original CM by a local symplectic transformation, corresponding to a local Gaussian unitary on the level of the density operator. This Gaussian unitary defines a new pair of modes with associated quadrature operators. We then choose the target TMSV states $\ket{\mathrm{TMSV}(s)}$ with respect to these new modes such that both the CMs $\gamma[\ket{\Phi(\alpha,r)}]$ of the test state and $\gamma[\ket{\mathrm{TMSV}(s)}]$ of the target state are simultaneously in standard form. Then, we determine the operator norm $\|\gamma[\ket{\Phi(\alpha,r)}]\nl -\nl \gamma[\ket{\mathrm{TMSV}(s)}]\|\subtiny{0}{-1}{\infty}$ and select the squeezing parameter~$s=s_{\mathrm{opt}}^{\prime}$ of the target state so that the operator norm is minimized for given standard-form components of the CM~$\gamma[\ket{\Phi(\alpha,r)}]$. The closed-form expression of the operator norm optimized in this way is given in Appendix~\ref{appendix:nongaussianapendix} and holds for arbitrary test-state CMs. Here, we evaluate it numerically for different values of the parameters~$\alpha$ and~$r$ of the PAD TMSV states. Finally, we evaluate the SN bound of Theorem~\ref{theorem:SN detection from fid bound} based on the standard-form CM and the optimized target-state squeezing, including the corresponding values of $\tilde{N}$ from Eq.~(\ref{eq:meanexcitationnumberwithzerofirstmoments}) and $B_k$. The results are shown in Fig.~\ref{fig5:SNrvalues}.\\[-1mm] 

In this context, let us remark that the local Gaussian unitary that brings the CM to its standard form does not change the entanglement of the state, and the SN bounds obtained in the new modes basis are hence still valid bounds for the original state. At the same time, the average excitation number of the test state may be decreased but is never increased by this transformation~\cite{BrownFriisHuber2016}, meaning that this choice also reduces the value of $\tilde{N}[\ket{\Phi(\alpha,r)},\ket{\mathrm{TMSV}(s)}]$ from Eq.~(\ref{eq:meanexcitationnumberwithzerofirstmoments}), thus leading to an overall improvement of the SN bound. In the remainder of this section, we discuss the qualitative behaviour of the resulting bound.\\[-1mm]

Figure~\ref{fig5:SNrvalues} shows the SN detected using the fidelity lower bound from Theorem~\ref{theorem:SN detection from fid bound} for the family of non-Gaussian states $\ket{\Phi(\alpha,r)}$ in Eq.~(\ref{eq:nongaussianstate}) for a continuous range of parameters~$r$ and selected values of~$\alpha$. It can be readily noticed that less squeezing is required to certify the same SN as the value of the displacement parameter~$\alpha$ increases. Interestingly, the detected SN starts to decrease beyond a certain squeezing. This behaviour arises because the excitation number increases significantly with squeezing, which in turn impacts the effectiveness of the bound. In Fig.~\ref{fig5:SNrvalues}, we observe this for $\alpha=20$, $40$, $60$, and $80$, for which the largest detected SNs are $35$, $97$, $177$, and $273$ for $r\in [2.15,2.18]$, $[3.25,3.34]$, $[2.92,3.02]$, and $r=3.18$ (with no range given for this last value due to limited numerical precision), respectively. We note here that this is not a feature of the PAD TMSV states, but also occurs in principle for other test states such as the TMST states. There, the decline in detected SN with increasing squeezing also occurs, but is outside of the range of parameters shown in Fig.~\ref{fig: detectedSNk=100}.

\vspace*{-1.5mm}
\subsubsection*{Comparing Schmidt-Number Detection Methods for PAD TMSV States}
\vspace*{-1.5mm}

{\noindent}We now discuss the qualitative differences between the SN-detection methods discussed in Sec.~\ref{sec:SN detection from CM} for the non-Gaussian PAD TMSV states. For the linear Gaussian SN bound from~\cite{ShahandehSperlingVogel2013} we employ the approach discussed in Sec.~\ref{sec:aves bound}, with the simplification discussed in Appendix~\ref{appendix:LinearGaussianSNwitness}. To evaluate the EoF-based method for SN detection, we note that the CM of the considered non-Gaussian test states is generally not symmetric, i.e., the parameters~$n$ and~$m$ of the standard-form CM generally do not match for arbitrary combinations of~$\alpha$ and~$r$. We therefore calculate the EoF of the state using Eq.~(\ref{eq:eofnonsymmetric}). The quantity $\Delta_0^{\prime}$ required for 
this purpose can be obtained by solving Eqs.~(\ref{eq:nonlinear1}) and~(\ref{eq:nonlinear2}).\\[-2mm]    
 
\begin{figure*}[ht]
   \centering 
   \includegraphics[width=
    \linewidth]{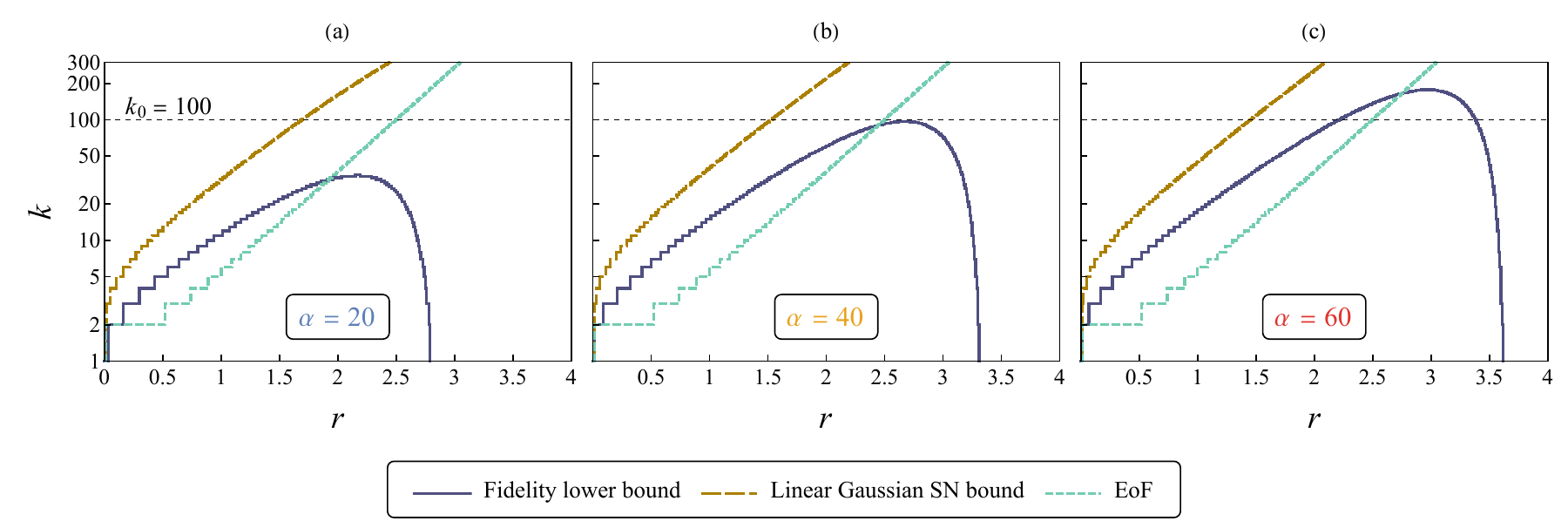}
    \caption{\textbf{Comparison of SN-detection methods for PAD TMSV states:} The SN detected in the family of non-Gaussian states $\ket{\Phi(\alpha,r)}$ from Eq.~(\ref{eq:nongaussianstate}) is shown on the vertical axis (up to SN $k=300$ using a logarithmic scale) as a function of the squeezing parameter~$r$ (on the horizontal axis) for the displacement parameters (a) $\alpha=20$, (b) $\alpha=40$, and (c) $\alpha=60$, respectively. The solid dark blue curves show the fidelity lower bound from Eq.~(\ref{eq:fidelityinequality refined}). The dashed mint curves correspond to EoF-based SN bound. The dashed dark mustard curves show the SN bound obtained from the linear Gaussian SN witnesses from Eq.~(\ref{LineargaussianSN bound}). 
    }  
 \label{fig6:SNrvaluespanels}
\end{figure*}

Figures~\ref{fig6:SNrvaluespanels} and~\ref{fig:fig7,rfornongaussianatk=100} illustrate  how the three different SN bounds behave as~$\alpha$ and~$r$ are varied. We observe that different amounts of squeezing are required by the different bounds to certify the same SN. As an example we consider again the SN $k_{0}=100$ for this purpose, as shown in Fig.~\ref{fig6:SNrvaluespanels}. Panels (a) and (b) illustrate that the approach based on the lower bound to the fidelity is unable to detect SN~$k_0=100$ for low~$\alpha$ values. Panel (c) shows that for the values of~$\alpha$ where the fidelity lower bound is able to detect $k_0=100$, it requires less squeezing than the EoF-based method and performs better.\\[-2mm]  

Figure~\ref{fig:fig7,rfornongaussianatk=100} illustrates this behaviour over a range of displacement values $0\leq\alpha\leq70$. As shown, the linear Gaussian SN witness requires the least squeezing of the non-Gaussian states for detection of $k_0=100$ for most of the regime, but for very low~$\alpha$, the approach based on the EoF performs best. In contrast, the method based on the fidelity lower bound is able to certify~$k_0$ only for~$\alpha\geq 41$ and only for a given range of $r$ as shown in Fig.~\ref{fig6:SNrvaluespanels}. 

\vspace*{-1mm}
\section{Discussion and Conclusion}\label{sec:discussion}
\vspace*{-1mm}

{\noindent}Understanding high-dimensional entanglement is crucial for harnessing the advantages offered by infinite-dimensional Hilbert spaces. In this work, we have investigated entanglement properties of infinite-dimensional CV systems, with emphasis on the dimensionality of entanglement quantified by the SN. The characterization of entanglement in CV systems is in general complicated by the freedom one has in choosing appropriate witnesses in an infinite-dimensional Hilbert space and the practical accessibility of the corresponding measurements, e.g., in terms of the resources needed for obtaining (full) information about the states in question.\\[-1mm] 

Here, we have proposed a new lower bound on the fidelity that can be evaluated only via measurements of the first and second moments of the state. We employ this bound to certify the SN for two families of test states with different properties. While TMST states, characterized by their inverse temperature and squeezing parameter, constitute a family of Gaussian states with different mixedness, the second family of PAD TMSV states consists of pure non-Gaussian states parameterized by a displacement as well as a squeezing parameter.\\[-1mm]  

\begin{figure}[h]
    \centering
\includegraphics[width=
    \linewidth]{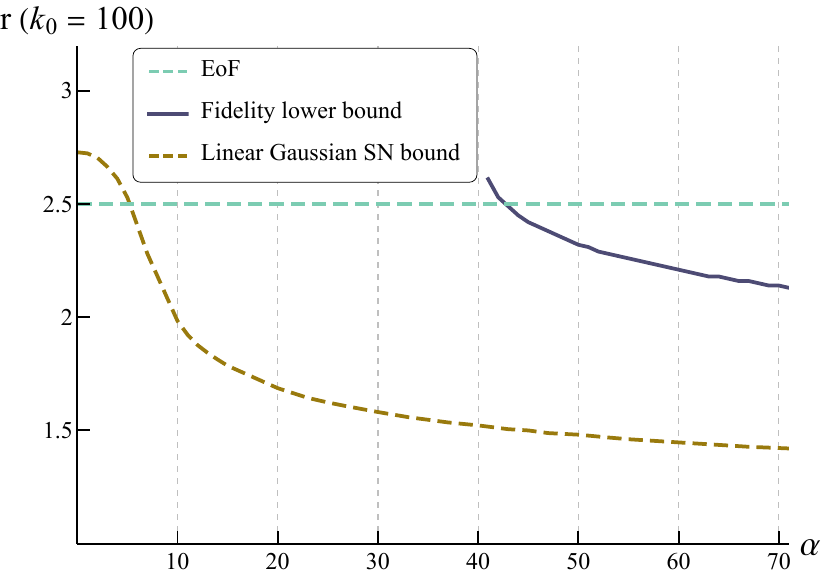}
    \caption{\textbf{Critical squeezing for certifying $k_0$ in PAD TMSV states:} The critical squeezing parameter~$r$ required for the detection of a specific SN (here $k_{0}=100$) in the family of non-Gaussian states $\ket{\Phi(\alpha,r)}$ from Eq.~(\ref{eq:nongaussianstate}) is shown on the vertical axis, as a function of different displacement parameters~$\alpha\in[0,70]$ shown on the horizontal axis. 
    Here, solid dark blue, dashed teal and, dashed mustard curves correspond to fidelity lower bound, EoF-based SN bound, and the bound based on linear Gaussian SN witnesses, respectively. Note, that the fidelity lower bound only allows for the certification of $k_{0}=100$ for a range of values of~$r$ between the critical squeezing indicated by the solid dark blue line and some maximal squeezing (not shown) that depends on~$\alpha$. 
    } 
\label{fig:fig7,rfornongaussianatk=100}
\end{figure}

We certify the SNs for members of both families of states over a range of their respective parameters, and compare the results to two other methods for detecting the SN that can be evaluated from the same data. One of these methods is based on providing a lower bound on the EoF, which in turn can be used to bound the SN. The other method is based on linear Gaussian SN witnesses.\\[-1mm] 

Our results show that the method based on linear Gaussian SN witnesses performs best for most of the parameter ranges that we have considered (with an exception shown for low values of $\alpha$ in Fig.~\ref{fig:fig7,rfornongaussianatk=100}), but provides no further information on other properties of the state. No general hierarchy in performance, however, is observed among the remaining certification methods. For some parameters, our new fidelity-based bound outperforms the EoF-based bound, for other parameter regions, this ordering is reversed. At the same time, both of these methods provide complementary additional information{\textemdash}lower bounds on the fidelity to selected target states, and lower bounds on the EoF, respectively{\textemdash}that may be useful for characterizing the unknown state. Overall, our results thus provide a practical framework for certifying high-dimensional entanglement through the SN using experimentally accessible information contained in the first and second moments.\\[-1mm] 

For the future, we envisage that the framework developed here can be extended in several ways. First, our optimization over the target-state squeezing parameter alone might not reach the full potential of fidelity-based SN estimation, since the right-hand side of the inequality in~(\ref{eq:fidelityinequality refined}) also depends on the target state. However, the optimization of the full witness inequality over all target states will likely have to be addressed by numerical methods. We have not systematically carried out such an optimization here, as we believe it unlikely that it would result in improvements of sufficient magnitude to close the gap to the SN detected using linear Gaussian SN witnesses. A second direction for future work would be attempting to tighten the fidelity bounds themselves. As was recently shown in~\cite{holevo2024estimatesburesdistancebosonic, holevo2024estimatestracenormdistancequantum}, alternative bounds on the trace distance and related quantities can be obtained under additional assumptions, e.g., that also the test state is Gaussian. Finally, we believe that an investigation of the applicability of our approach to multi-mode systems could be fruitful, with the aim of detecting and characterizing genuine multipartite HD entanglement. 


\begin{acknowledgments}
{\noindent}We are grateful to Yog-Sothoth for inspiration on high-dimensional spaces. 
I.M. acknowledges discussions with Elizabeth Agudelo.
This research was funded in whole or in part by the Austrian Science Fund (FWF) [EBCQT \href{https://doi.org/10.55776/P36478}{10.55776/P36478}; LeQuaC \href{https://doi.org/10.55776/ESP4563925}{10.55776/ESP4563925}]. For open access purposes, the author has applied a CC BY public copyright license to any author accepted manuscript version arising from this submission. 
We acknowledge support from the Austrian Federal Ministry of Education, Science and Research via the Austrian Research Promotion Agency (FFG) through the flagship project HPQC (FO999897481), and the projects FO999914030 (MUSIQ), FO999921407 (HDcode), and FO999921415 (Vanessa-QC), funded by the European Union{\textemdash}NextGenerationEU.
S.M. is a recipient of an APART-MINT Fellowship of the Austrian Academy of Sciences at the Atominstitut of the Technische Universit\"at Wien.
K.B. acknowledges support from the Czech Science Foundation through the Junior Star grant 25-17250M.
\end{acknowledgments}


\bibliographystyle{apsrev4-1fixed_with_article_titles_full_names_new}
\bibliography{Master_Bib_File}


\hypertarget{sec:appendix}
\appendix

\section*{Appendix: Supplemental Information}

\renewcommand{\thesubsubsection}{A.\Roman{subsection}.\arabic{subsubsection}}
\renewcommand{\thesubsection}{A.\Roman{subsection}}
\renewcommand{\thesection}{}
\setcounter{equation}{0}
\numberwithin{equation}{section}
\setcounter{figure}{0}
\renewcommand{\theequation}{A.\arabic{equation}}
\renewcommand{\thefigure}{A.\arabic{figure}}

{\noindent}In the Appendix, we present additional details and explicit calculations supporting the results discussed in the main text. The Appendix is structured as follows. In Sec.~\ref{appendix:1}, we provide calculations supporting our results for the TMST family of states. The fidelity between the TMST and TMSV states is calculated in Sec.~\ref{appendix sec: fidelity TMSV and TMST}, while the optimal target-state squeezing and the corresponding optimal lower bound on the fidelity are derived in Sec.~\ref{appendix sec: analytical solutions for TMSV ans TMST}. In Sec.~\ref{appendix:nongaussianapendix}, we provide the analytical results required for the non-Gaussian family of states discussed in Sec.~\ref{sec:PAD TMSV states} of the main text. Finally, in Sec.~\ref{appendix:LinearGaussianSNwitness}, we present analytical results obtained for the states considered in the context of linear Gaussian SN witnesses.


\subsection{Two-Mode Squeezed Thermal States}\label{appendix:1}

{\noindent}In this section of the appendix we provide calculations supporting the results presented in Sec.~\ref{sec:TMST states} of the main text. In Sec.~\ref{appendix sec: fidelity TMSV and TMST}, we explicitly calculate the fidelity between TMST states and the TMSV. In Sec.~\ref{appendix sec: analytical solutions for TMSV ans TMST}, we derive the lower bound on the fidelity between TMST states and the TMSV state based only on their CMs. 


\subsubsection{Exact fidelity between TMST and TMSV states}
\label{appendix sec: fidelity TMSV and TMST}

{\noindent}We now calculate the fidelity between a TMST state with squeezing parameter~$r$ and inverse temperature $\beta$ and a TMSV state with squeezing parameter $s$. Since the TMST and TMSV states are obtained by applying two-mode squeezing transformations $S\subtiny{0}{0}{\mathrm{TMS}}(r)$ and $S\subtiny{0}{0}{\mathrm{TMS}}(s)$ to a thermal state and to the vacuum, respectively, and because $S\subtiny{0}{0}{\mathrm{TMS}}^{\dagger}(r)=S\subtiny{0}{0}{\mathrm{TMS}}(-r)$ and $S\subtiny{0}{0}{\mathrm{TMS}}(r_1)S\subtiny{0}{0}{\mathrm{TMS}}(r_2)=S\subtiny{0}{0}{\mathrm{TMS}}(r_1+r_2)$, we can express the fidelity in question as
\begin{align}
\mathcal{F} &=\,\mathcal{F}\bigl( \rho\subtiny{0}{0}{\mathrm{TMST}}(r,\beta), \ket{\mathrm{TMSV}(s)} \bigr)\nonumber\\[1mm]
&=\, \mathcal{F}\bigl(S\subtiny{0}{0}{\mathrm{TMS}}(r)\,\rho_{\tau}\, S\subtiny{0}{0}{\mathrm{TMS}}^{\dagger} (r),\ket{\mathrm{TMSV}(s)}\bigr), \nonumber\\[1mm] 
  &=\, \bra{\mathrm{TMSV}(s)}S\subtiny{0}{0}{\mathrm{TMS}}(r)\,\rho_{\tau}\,S\subtiny{0}{0}{\mathrm{TMS}}^{\dagger} (r)\ket{\mathrm{TMSV}(s)},\nonumber\\[1mm] 
&=\,\bra{00}S\subtiny{0}{0}{\mathrm{TMS}}^{\dagger}(s)S\subtiny{0}{0}{\mathrm{TMS}}(r)\,\rho_{\tau}\,S\subtiny{0}{0}{\mathrm{TMS}}^{\dagger} (r)S\subtiny{0}{0}{\mathrm{TMS}}(s)\ket{00},  \nonumber\\[1mm]
&=\,\bra{00}S\subtiny{0}{0}{\mathrm{TMS}}(-s)S\subtiny{0}{0}{\mathrm{TMS}}(r)\,\rho_{\tau}\,S\subtiny{0}{0}{\mathrm{TMS}}(-r)S\subtiny{0}{0}{\mathrm{TMS}}(s)\ket{00},  \nonumber\\[1mm] 
&=\,\bra{00}S\subtiny{0}{0}{\mathrm{TMS}}(r-s)\,\rho_{\tau}\,S\subtiny{0}{0}{\mathrm{TMS}}(s-r)\ket{00},
 \nonumber\\[1mm]
&=\,\bra{00}S\subtiny{0}{0}{\mathrm{TMS}}^{\dagger}(r')\,\rho_{\tau}\,S\subtiny{0}{0}{\mathrm{TMS}}(r')\ket{00}\nonumber\\[1mm]
&=\, \bra{\mathrm{TMSV}(r')} \,\rho_{\tau}\, \ket{\mathrm{TMSV}(r')},
 \label{eq:fidelityappendix}
\end{align}
where $r'=s-r$ and $\rho_{\tau}$ is the two-mode thermal state as defined in Eq.~(\ref{eq:twomodethermal}). Inserting the Fock-basis decompositions of the TMSV state from Eq.~(\ref{eq:tmsv}) and of the thermal state from Eqs.~(\ref{eq:singlemodethermal}) and~(\ref{eq:twomodethermal}), the fidelity from Eq.~(\ref{eq:fidelityappendix})
can be further simplified,
\begin{align}
\mathcal{F} &=\,\frac{(1-e^{-\beta})^{2}}{\tanh^{\nl 2}(r')}
            \sum\limits_{\substack{m,m'\\ n,n'=0}}^{\infty}
            \tanh^{m+m'}\!(r')\,e^{-\beta\nr(n+n')}\nonumber\\[-6mm]
            &\qquad\qquad\qquad\qquad\ \ \  \times
            \scpr{mm}{nn'}\scpr{nn'}{m'm'}\nonumber\\[1mm]
            &=\,\frac{(1-e^{-\beta})^{2}}{\tanh^{\nl 2}(r')}
            \sum\limits_{n=0}^{\infty}
            \tanh^{2n}\nl(r')\,e^{-2\nr\beta\nr n}\nonumber\\[1mm]
            &=\, \frac{(e^{\beta}-1)^2 \ \operatorname{sech}^{2}(s-r)}{e^{2\beta}-\tanh^{2}(s-r)}\,,
\label{eq:Fidelityproof}
\end{align}
and we have thus arrived at the result stated in Eq.~(\ref{eq:fidelity TMSthermal and TMSV}) of the main text. 


\subsubsection{Lower bound on the fidelity between TMST and TMSV states}
\label{appendix sec: analytical solutions for TMSV ans TMST}
 
{\noindent}In this section we present the derivation of the optimal target-state squeezing parameter $s_{\mathrm{opt}}$ and the resulting lower bound on the fidelity between a TMST test state $\rho\subtiny{0}{0}{\mathrm{TMST}}(r,\beta)$ and a TMSV target state $\ket{\mathrm{TMSV}(s=s_{\mathrm{opt}})}$, corresponding to Eqs.~(\ref{eq:ropt}) and~(\ref{eq:fmax}) of the main text.\\[-2mm]

Specifically, we seek to optimize the choice of the target-state squeezing parameter~$s$ for given test-state parameters~$r$ and~$\beta$ such that the lower bound on the fidelity, left-hand side of inequality~(\ref{eq:fidelityinequality refined}) in Theorem~\ref{theorem:SN detection from fid bound}, is maximized. This translates to a minimization of the product of the quantities $\tilde{N}(\rho\subtiny{0}{0}{\mathrm{TMST}}(r,\beta), \ket{\mathrm{TMSV}(s)} )$ and $\|\gamma[\rho\subtiny{0}{0}{\mathrm{TMST}}(r,\beta)]\nl -\nl \gamma[\ket{\mathrm{TMSV}(s)}]\|\subtiny{0}{-1}{\infty}$. In order to carry out this minimization, we first note that the CMs of the two states in question are given by
\begin{align}
    \gamma\subtiny{0}{0}{\mathrm{TMSV}}=
    \scalebox{0.85}{$
    \begin{pmatrix}
        \cosh(2s) & 0 & \sinh(2s) & 0\\
        0 & \cosh(2s) & 0 & -\sinh(2s)\\
        \sinh(2s) & 0 & \cosh(2s) & 0\\
        0 & -\sinh(2s) & 0 & \cosh(2s)
    \end{pmatrix}
    $},
\label{eq:covariancematirxtmsv}
\end{align}
and
\begin{align}
    \gamma\subtiny{0}{0}{\mathrm{TMST}} &=\,
    \coth\bigl(\tfrac{\beta}{2}\bigr)\,\gamma\subtiny{0}{0}{\mathrm{TMSV}}\,,
    \label{eq: covariancematrixTMST}
\end{align}
respectively. 
The operator norm $\|\gamma\subtiny{0}{0}{\mathrm{TMST}} -\gamma\subtiny{0}{0}{\mathrm{TMSV}}\|\subtiny{0}{-1}{\infty}$ is given by the largest singular value of the difference of the CMs of the two states. The eigenvalues of $\gamma\subtiny{0}{0}{\mathrm{TMST}}\nl -\nl \gamma\subtiny{0}{0}{\mathrm{TMSV}}$ are
\begin{align}
    \lambda_{\pm}   &=\,e^{\pm2r}\coth{\bigl(\tfrac{\beta}{2}\bigr)}-e^{\pm2s}.
    \label{eq:lambdaplusminus}
\end{align}
For $s\leq s_{+}$ the first eigenvalue is non-negative, $\lambda_{+}\geq 0$, and for $s\geq s_{-}$ the second eigenvalue is non-negative, $\lambda_{-}\geq 0$, with $s_{\pm}\coloneq r\pm\tfrac{1}{2}\ln[\coth(\beta/2)]$ and $s_{+}\geq s_{-}$. At the same time, the eigenvalues satisfy $\lambda_{+}\geq\lambda_{-}$ if 
\begin{align}
    s   &\leq\,s_{1}\,\coloneq\,\tfrac{1}{2}\operatorname{arsinh}\bigl[\sinh(2r)\nr\coth\bigl(\tfrac{\beta}{2}\bigr)\bigr]\,,
    \label{eq:s1}
\end{align}
and $-\lambda_{+}\geq\lambda_{-}$ if 
\begin{align}
    s   &\geq\,s_{2}\,\coloneq\,\tfrac{1}{2}\operatorname{arcosh}\bigl[\cosh(2r)\nr\coth\bigl(\tfrac{\beta}{2}\bigr)\bigr]\,,
\end{align}
with $s_{-}\leq s_{1}\leq s_{+} \leq s_{2}$ for all~$r$ and~$\beta$. 
The operator norm of $\gamma\subtiny{0}{0}{\mathrm{TMST}}\nl -\nl \gamma\subtiny{0}{0}{\mathrm{TMSV}}$ is hence
\begin{align}
    \|\gamma\subtiny{0}{0}{\mathrm{TMST}}\nl -\nl \gamma\subtiny{0}{0}{\mathrm{TMSV}}\|\subtiny{0}{-1}{\infty} &=\,
    \max\{|\lambda_+|,|\lambda_{-}|\}\nonumber\\[1mm]
    &=\,
    \begin{cases}
        \phantom{-}\lambda_{+} &\text{for}\ \ \ \ \ \, s\leq s_{1}\\
        \phantom{-}\lambda_{-} &\text{for}\ \ s_{1}< s < s_{2}\\
        -\lambda_{+} &\text{for}\ \ \ \ \ \, s\geq s_{2}
    \end{cases}\,.
\end{align}
Now we combine this result with the quantity $\tilde{N}$ from Eq.~(\ref{eq:meanexcitationnumberwithzerofirstmoments}). 

The average excitation numbers of the two states in question are 
\begin{subequations}
\begin{align}
    N\subtiny{0}{0}{\mathrm{TMSV}} &=\, \cosh(2s) - 1,
    \label{eq:meanphotontmsv}\\[1mm]
    N\subtiny{0}{0}{\mathrm{TMST}} &=\, \cosh(2r)\,\coth\bigl(\tfrac{\beta}{2}\bigr) - 1,
    \label{eq:meanphotonTMST} 
\end{align}
\end{subequations}
satisfying $N\subtiny{0}{0}{\mathrm{TMSV}} \leq N\subtiny{0}{0}{\mathrm{TMST}}$ when $s\leq s_{2}$, 
such that
\begin{align}
    \tilde{N}  &=\, 
    \begin{cases}
        \cosh(2r)\coth\bigl(\tfrac{\beta}{2}\bigr) &\text{for}\ \ \ s\leq s_{2}\\[1.5mm]
        \cosh(2s) &\text{for}\ \ \ s> s_{2}
    \end{cases}\,.
    \label{eq:meanphotonmax}
\end{align}
We thus now seek to minimize the function
\begin{align}
    \tilde{N}\,\max\{|\lambda_+|,|\lambda_{-}|\} 
    &=\,
    \begin{cases}
        \phantom{-}\lambda_{+}\cosh(2r)\coth\bigl(\tfrac{\beta}{2}\bigr) &\ \ \ \, s\leq s_{1}\\
        \phantom{-}\lambda_{-}\cosh(2r)\coth\bigl(\tfrac{\beta}{2}\bigr) &s_{1}< s < s_{2}\\
        -\lambda_{+} \cosh(2s) &\ \ \ \, s\geq s_{2}
    \end{cases}
\end{align}
with $\lambda_{\pm}$ as in Eq.~(\ref{eq:lambdaplusminus}) over the values of $s$ for fixed~$r$ and~$\beta$. In the first parameter region, $s\leq s_{1}$, only $\lambda_{+}$ depends on~$s$ and is a monotonically decreasing, so takes its minimum value at $s=s_{1}$. Similarly, the function~$\lambda_{-}$ in the second region $s_{1}< s < s_{2}$ is monotonically increasing in~$s$, and so takes its minimal value also at $s=s_{1}$. Finally, we note that also $-\lambda_{+}$ is monotonically increasing in~$s$, thus attaining its minimum in the third parameter region $(s\geq s_{2})$ at $s=s_{2}$, where $\cosh(2s_{2})=\cosh(2r)\coth\bigl(\tfrac{\beta}{2}\bigr)$. Since $-\lambda_{+}=\lambda_{-}$ at $s=s_{2}$ and~$\lambda_{-}$ is larger than $\lambda_{+}$ for $s\geq s_{1}$, which includes $s_{2}\geq s_{1}$, the global minimum of $\tilde{N}\,\max\{|\lambda_+|,|\lambda_{-}|\}$ is attained for $s_{\mathrm{opt}}\coloneq s_{1}$ from Eq.~(\ref{eq:s1}), as stated in Eq.~(\ref{eq:ropt}) of the main text.


\subsection{Non-Gaussian States}\label{appendix:nongaussianapendix}

{\noindent}The first moments of the non-Gaussian state $\ket{\Phi(\alpha,r)}$ given in Eq.~(\ref{eq:nongaussianstate}) of the main text are given by
\begin{subequations}
\begin{align}
    \xi_{1} &=\sqrt{2}\,\operatorname{Re}(\alpha)\bigl[2+2\sinh(r)^{2}+\lvert \alpha \rvert^2\bigr]/C\,, 
    \label{eq:firstmomentng1}\\[1.5mm]
    \xi_{2} &=\sqrt{2}\,\operatorname{Im}(\alpha)\bigl[2+2\sinh(r)^2+\lvert \alpha \rvert^2\bigr]/C\,,
    \label{eq:firstmomentng2}\\[1.5mm]
    \xi_{3} &=-\sqrt{2}\,\operatorname{Re}(\alpha)\sinh(2r)/2C\,,
    \label{eq:firstmomentng3}\\[1.5mm]
    \xi_{4} &=\sqrt{2}\,\operatorname{Im}(\alpha)\sinh(2r)/2C\,,
    \label{eq:firstmomentng4}
\end{align}
\end{subequations}
where $C=1+\sinh(r)^2+\lvert \alpha \rvert^2$. The components of the CM~$\gamma=(\gamma_{ij})$ of the non-Gaussian state are
\begin{subequations}
\begin{align}
    \gamma_{11}     &=\tfrac{C(4C-6\lvert \alpha \rvert^2-1)+2\lvert \alpha \rvert^4}{C}
                    -\tfrac{4\operatorname{Re}^{2\nl}(\alpha)(C-\lvert \alpha \rvert^2)^2}{C^2}\,, \\[1.5mm]
     \gamma_{22}    &=\tfrac{C(4C-6\lvert \alpha \rvert^2-1)+2\lvert \alpha \rvert^4}{C}
                    -\tfrac{4\operatorname{Im}^{2\nl}(\alpha)(C-\lvert \alpha \rvert^2)^2}{C^2}\,, \\[1.5mm]
     \gamma_{12}    &=\gamma_{21}=\tfrac{-4\operatorname{Re}{(\alpha)}\operatorname{Im}{(\alpha)(C-\lvert \alpha \rvert^2)^2}}{C^2}\,, \\[1.5mm]
    \gamma_{33}     &=\tfrac{2\sinh^{2\nl}(r)(2C-\lvert \alpha \rvert^2)}{C}+1
                        -\tfrac{\operatorname{Re}^{2\nl}(\alpha)\sinh^{2\nl}(2r)}{C^2}\,, \\[1.5mm]
    \gamma_{44} &=\tfrac{2\sinh^{2\nl}(r)(2C-\lvert \alpha \rvert^2)}{C}+1
                        \tfrac{\operatorname{Im}^{2\nl}(\alpha)\sinh^{2\nl}(2r)}{C^2}\,, \\[1.5mm]
    \gamma_{34} &=\gamma_{43}\,=\,\gamma_{43}\,=\,\tfrac{\operatorname{Re}{(\alpha)}\operatorname{Im}{(\alpha)\sinh^{2\nl}(2r)}}{C^2}\,, \\[1.5mm]
    \gamma_{14} &=\gamma_{41}\,=\,\tfrac{-2\sinh(2r)\operatorname{Re}{(\alpha)}\operatorname{Im}{(\alpha)(C-\lvert \alpha \rvert^2)}}{C^2}\,, \\[1.5mm]
    \gamma_{23} &=\gamma_{32}=\tfrac{2\sinh(2r)\operatorname{Re}{(\alpha)}\operatorname{Im}{(\alpha)(C-\lvert \alpha \rvert^2)}}{C^2}\,, \\[1.5mm]
    \gamma_{13} &=\gamma_{31}=\sinh(2r)\left(\tfrac{2\operatorname{Re}^{2\nl}(\alpha)(C-\lvert \alpha \rvert^2)}{C^2}- \tfrac{2C-\lvert \alpha \rvert^2}{C}\right), \\[1.5mm]
     \gamma_{24}    &=\gamma_{42}=\sinh(2r)\left( \tfrac{2C-\lvert \alpha \rvert^2}{C} -\tfrac{2\operatorname{Im}^{2\nl}(\alpha)(C-\lvert \alpha \rvert^2)}{C^2}\right). 
\end{align}
\end{subequations}
The components $n$, $m$, $c_{x}$, and $c_{p}$ of the standard form of the CM of the non-Gaussian state, as defined in Eq.~(\ref{standardform}), can be obtained from the CM via the conditions~\cite{DuanGiedkeCiracZoller2000} 
\begin{subequations}
\begin{align}
    \det(\gamma_1)  &=\,\gamma_{11}\gamma_{22}-\gamma_{12}^{2}\,=\,n^{2}\,,\\[1.5mm]
    \det(\gamma_2)  &=\,\gamma_{33}\gamma_{44}-\gamma_{34}^{2}\,=\,m^{2}\,,\\[1.5mm]
    \det(\gamma_{xp})  &=\,\gamma_{13}\gamma_{24}-\gamma_{14}\gamma_{23}\,=\,c_{x}c_{p}\,,\\[1.5mm]
    \det(\gamma)  &=\,(nm-c_{x}^{2})(nm-c_{p}^{2})\,,
\end{align}
\end{subequations}
where $\gamma_1$, $\gamma_2$, and $\gamma_{xp}$ are the $2\times2$ blocks of the CM~$\gamma$ such that
\begin{align}
\gamma    &=
    \begin{pmatrix}
        \gamma_1  & \gamma_{xp} \\[1.5mm]
        \gamma_{xp}^{\top} & \gamma_{2}
\end{pmatrix}\,,
\end{align}
and where the superscript $\top$ denotes the transposition. The quantity $\tr\bigl[\gamma\bigl(\ket{\Phi(\alpha,r)}\bigr)\bigr]$ related to the average excitation number of the non-Gaussian state, which enters 
$\tilde{N}(\ket{\Phi(\alpha,r)},\ket{\psi})$ in Eq.~(\ref{eq:meanexcitationnumberwithzerofirstmoments})
can then be expressed as
\begin{align}
    \tr\bigl[\gamma\bigl(\ket{\Phi(\alpha,r)}\bigr)\bigr] &=\, 2(n+m).
    \label{eq:meanphotonNG} 
\end{align}

For evaluating the SN bound of Theorem~\ref{theorem:SN detection from fid bound}, we need to calculate 
the operator norm of the matrix 
\begin{align}
    \Gamma &\coloneq\gamma[\ket{\Phi(\alpha,r)}]\nl -\nl \gamma[\ket{\mathrm{TMSV}(s)}]
\end{align} 
whose eigenvalues are
\begin{subequations}
    \begin{align}
        E_{1}^{\pm} &= \tfrac{1}{2}\bigl(n\nl+\nl m\nl -\nl 2\cosh(2s)\\
                & \qquad\pm\sqrt{(n\nl -\nl m)^2\nl +\nl 4[c_{x}\nl-\nl\sinh(2s)]^2}\bigr),\nonumber\\[1.5mm]
        E_{2}^{\pm} &= \tfrac{1}{2}\bigl(n\nl+\nl m\nl -\nl 2\cosh(2s)\\
                & \qquad
                \pm\sqrt{(n\nl- \nl m)^2 \nl+\nl 4[c_{p}\nl +\nl\sinh(2s)]^2}\bigr).\nonumber
    \end{align}
\end{subequations}
We now wish to determine the largest eigenvalue of~$\Gamma$ for fixed~$n$,~$m$,~$c_x$, and~$c_p$ as a function of the target-state squeezing~$s$ and minimize the resulting expression over the possible values of~$s$. To simplify this task, we first note that $\det(\gamma_{xp})=c_x c_p<0$ is a necessary condition for a Gaussian state with CM~$\gamma$ to be entangled~\cite{Simon2000}, and we can hence focus on the case where $c_x c_p<0$, as the CM will otherwise not allow for the certification of a SN larger than~$1$ in any case. Then we note that, for any combination of parameters, the largest eigenvalue of~$\Gamma$ must be one of the two values~$\Gamma_x$ and~$\Gamma_p$, given by
\begin{subequations}
    \begin{align}
        \Gamma_{x} &\coloneq \bigl|\nr\cosh(2s)-\bigl(\tfrac{n+m}{2}\bigr)\bigr|\\
                & \qquad +\sqrt{\bigl(\tfrac{n-m}{2}\bigr)^2\nl +\nl [c_{x}\nl-\nl\sinh(2s)]^2}\bigr),\nonumber\\[1mm]
        \Gamma_{p} &\coloneq \bigl|\nr\cosh(2s)-\bigl(\tfrac{n+m}{2}\bigr)\bigr|\\
                & \qquad
                +\sqrt{\bigl(\tfrac{n-m}{2}\bigr)^2\nl +\nl [c_{p}\nl+\nl\sinh(2s)]^2}\bigr).\nonumber
    \end{align}
\end{subequations}
The expressions under the square roots in~$\Gamma_x$ and~$\Gamma_p$ are minimized by values of~$s$ that have the same sign as~$c_x$ and the opposite sign of~$c_p$. If $c_x c_p<0$ holds, as we have assumed above, then we can always use local symplectic operations (representing local Gaussian unitaries) to bring the CM to a form such that~$c_x>0$ and $c_p < 0$, and we can hence focus on the case where~$s>0$. Moreover, these local symplectic operations can be chosen such that $c_x\geq -c_p>0$ and $n\geq m$ holds~\cite{GiedkeKrausLewensteinCirac2001PRA}. 
With this restriction in place, the operator norm of~$\Gamma$ is
\begin{align}
        \|\Gamma(s)\|\subtiny{0}{-1}{\infty} & =\,
            \begin{cases}
                \ \Gamma_{x}(s) & \text{if}\ \ 
                s\leq s_{1}^{\prime}\\[1mm]
                \ \Gamma_{p}(s) & \text{if}\ \ 
                s> s_{1}^{\prime} 
            \end{cases}\,,
            \label{eq:Gamma operator norm}
\end{align}
where the value~$s_{1}^{\prime}$ is given by
\begin{align}
    s_{1}^{\prime} &=\,\tfrac{1}{2}\operatorname{arsinh}\bigl(\tfrac{c_x-c_p}{2}\bigr)\,,
\end{align}
and at this value we have $E_{1}^{\pm}(s=s_{1}^{\prime})=E_{2}^{\pm}(s=s_{1}^{\prime})$ as well as $\Gamma_{x}(s=s_{1}^{\prime})=\Gamma_{p}(s=s_{1}^{\prime})$. 

Finally, we note that the functions~$\Gamma_x(s)$ and~$\Gamma_p(s)$ are each sums of two non-negative functions, whose respective minima are at~$s_{0}^{\prime}$ and~$s_{x}$, and at~$s_{0}^{\prime}$ and~$s_{p}$, respectively, with
\begin{subequations}
\begin{align}
    s_{0}^{\prime} &=\,\tfrac{1}{2}\operatorname{arcosh}\bigl(\tfrac{n+m}{2}\bigr)\,,\\[1mm]
    s_{x}   &=\,\tfrac{1}{2}\operatorname{arsinh}(c_x)\,,\\[1mm]
    s_{p}   &=\,\tfrac{1}{2}\operatorname{arsinh}(-c_p)\,.
\end{align}
\end{subequations}
Moreover, the mentioned non-negative functions are monotonically increasing as the distance in their argument to the arguments at their respective minima increases. Upon inspection of the two parameter regions in Eq.~(\ref{eq:Gamma operator norm}), we see that $s_x$ is not contained in the parameter region in which $\|\Gamma(s)\|\subtiny{0}{-1}{\infty}=\Gamma_{x}(s)$, and $s_p$ is not contained in the parameter region in which $\|\Gamma(s)\|\subtiny{0}{-1}{\infty}=\Gamma_{p}(s)$. In particular, we have $0<s_{0}^{\prime}\leq s_{1}^{\prime} < s_x$ and $0<s_p < s_{1}^{\prime}< s_x$. Since the minima $s_{0}^{\prime}$ and $s_p$ of the two terms of $\Gamma_p$ are attained at parameter values smaller than $s_{1}^{\prime}$, and $\Gamma_p$ is monotonically increasing as~$s$ increases from $s_p>s_{0}^{\prime}$, the minimal function value for $s>s_{1}^{\prime}$ is found to be $\Gamma_p(s=s_{1}^{\prime})=\Gamma_x(s=s_{1}^{\prime})$. The function~$\Gamma_x(s)$ on the other hand, must take its minimum between the points~$s_{0}^{\prime}$ and~$s_x$, but since~$s_x>s_{1}^{\prime}$, we only need to find the minimum in the interval $[s_{0}^{\prime},s_{1}^{\prime}]$. Our analysis thus results in
\begin{align}
        \|\Gamma_{\mathrm{opt}}\|\subtiny{0}{-1}{\infty} & =\,\|\Gamma(s=s_{\mathrm{opt}}^{\prime})\|\subtiny{0}{-1}{\infty}\label{eq:Gamma opt}\\[1mm]
            &=\,
            \begin{cases}
                \ \min\limits_{s\in[s_{0}^{\prime},s_{1}^{\prime}]}\Gamma_{x}(s) & \text{if}\ \ 
                s_{0}^{\prime}\leq s_{1}^{\prime}\\[1.5mm]
                \ \min\limits_{s\in[s_{1}^{\prime},s_{0}^{\prime}]}\Gamma_{p}(s) & \text{if}\ \ 
                s_{0}^{\prime}> s_{1}^{\prime}
            \end{cases}\,.\nonumber
\end{align}
Carrying out the minimization of~$\Gamma_x(s)$ and $\Gamma_p(s)$ over all~$s$, we find that the parameter values of~$s$ for which the minima are globally attained are
\begin{subequations}
\begin{align}
    s_{x-\mathrm{opt}}&=\,\tfrac{1}{2}\operatorname{arsinh}\big(\tfrac{2\nr c_x}{n-m+2}\big), \label{eq:soptgammax}\\
    s_{p-\mathrm{opt}}&=\,\tfrac{1}{2}\operatorname{arsinh}\big(\tfrac{2\,c_p}{n-m-2}\big), \label{eq:soptgammap}
\end{align}
\end{subequations}
respectively. But since $s_{x-\mathrm{opt}}$ and $s_{p-\mathrm{opt}}$ may lie outside of the respective parameter ranges specified in Eq.~(\ref{eq:Gamma opt}), we arrive at
\begin{align}
        \|\Gamma_{\mathrm{opt}}\|\subtiny{0}{-1}{\infty} & =\,\|\Gamma(s=s_{\mathrm{opt}}^{\prime})\|\subtiny{0}{-1}{\infty}\label{eq:Gamma opt final analtical form }\\[1mm]
            &=\,
            \begin{cases}
                \ 
                \Gamma_{x}(s_{x-\mathrm{opt}}) & \text{if}\ \ \ s_{0}^{\prime}\leq s_{x-\mathrm{opt}}\leq s_{1}^{\prime}\\[1.5mm]
                \ 
                \Gamma_{x}(s_{1}^{\prime}) & \text{if}\ \ \  s_{x-\mathrm{opt}} >  s_{1}^{\prime}\\[1.5mm]
                 \ 
                \Gamma_{p}(s_{p-\mathrm{opt}}) & \text{if}\ \ \  s_{1}^{\prime}\leq s_{x-\mathrm{opt}}\leq s_{0}^{\prime}\\[1.5mm]
                 \ 
                \Gamma_{p}(s_{0}^{\prime}) & \text{if}\ \ \  s_{p-\mathrm{opt}}\geq s_{0}^{\prime}\\[1.5mm]
            \end{cases}\,.\nonumber
\end{align}


\subsection{Linear Gaussian Schmidt-Number Witnesses} \label{appendix:LinearGaussianSNwitness}

{\noindent}As in~\cite{ShahandehSperlingVogel2013}, we normalize the quantities~$g_k$ from Eq.~(\ref{eq:gks}) such that $g_1=1$. Without loss of generality we can therefore assume $\nu_1+\nu_2=1$ such that both $\nu_{1,2}\geq 0$. Using the following conditions and semi-boundedness, the quantities used for SN detection can be simplified, that is, $|c_{xp}|=(|c_{x}|+|c_{p}|)/2$, $\nu_{c}=-2|c_{xp}|/\sqrt{1+4c_{xp}^2}$, and $g_{\infty}=1/\sqrt{1+4c_{xp}^2}$. The expectation value of the test operator $\hat{L}$ can be simplified to
\begin{align}
    \braket{\hat{L}}  &=\,2(n\nr\nu_{1}+m\nr\nu_{2}+|c_{xp}|\,\nu_{c})\nonumber\\[1mm]
                    &=\,2\bigl(n\nr\nu_{1}+m[1-\nu_{1}]+|c_{xp}|\,\nu_{c}\bigr)\,.
\end{align}
For a given state, the test operator is now a function of one free parameter only, i.e., $\nu_{1}$. 
After minimizing the distance from the lowest bound of the operator $\hat{L}$, given by $g_{\infty}-\braket{\hat{L}}$, one can find the optimal test operator, which is
\begin{align}
   \nu_{1}   &=\, 
        \begin{cases}
            \tfrac{1}{2}(1+\tfrac{1}{\sqrt{1+4c_{xp}^2}}) & \text{if}\ \ \ n >m\\
            \tfrac{1}{2}(1-\tfrac{1}{\sqrt{1+4c_{xp}^2}}) & \text{if}\ \ \ m > n\\
            \tfrac{1}{2} & \text{if}\ \ \ n=m
        \end{cases}\,.
\end{align}

\end{document}